\documentclass[longauth]{aa}

\usepackage{graphicx}
\usepackage{natbib}
\usepackage{scalerel}
\usepackage{comment}
\usepackage{booktabs}

\usepackage{placeins}

\usepackage[table]{xcolor}

\usepackage{subcaption}
\usepackage[font=small,labelfont=bf]{caption}

\usepackage{txfonts}

\usepackage[pdfencoding=auto,psdextra]{hyperref}
\hypersetup{
    colorlinks=true,
    linkcolor=blue,
    filecolor=magenta,      
    urlcolor=blue,
    citecolor=blue
}
\makeatletter
\renewcommand*\aa@pageof{, page \thepage{} of \pageref*{LastPage}}
\makeatother

\usepackage[utf8]{inputenc}

\usepackage[switch, modulo]{lineno}

\usepackage{euclid}

\begin{document}

\title{\Euclid preparation} \subtitle{LXXIX. Using mock low surface brightness dwarf galaxies to probe Euclid Wide Survey detection capabilities}

\newcommand{\orcid}[1]{}                   
\author{Euclid Collaboration: M.~Urbano\orcid{0000-0001-5640-0650}\thanks{\email{mathias.urbano@gmail.com}}\inst{\ref{aff1}}
\and P.-A.~Duc\orcid{0000-0003-3343-6284}\inst{\ref{aff1}}
\and M.~Poulain\orcid{0000-0002-7664-4510}\inst{\ref{aff2}}
\and A.~A.~Nucita\inst{\ref{aff3},\ref{aff4},\ref{aff5}}
\and A.~Venhola\orcid{0000-0001-6071-4564}\inst{\ref{aff2}}
\and O.~Marchal\orcid{0000-0001-7461-8928}\inst{\ref{aff1}}
\and M.~K\"ummel\orcid{0000-0003-2791-2117}\inst{\ref{aff6}}
\and H.~Kong\orcid{0000-0001-8731-1212}\inst{\ref{aff7}}
\and F.~Soldano\inst{\ref{aff8}}
\and E.~Romelli\orcid{0000-0003-3069-9222}\inst{\ref{aff9}}
\and M.~Walmsley\orcid{0000-0002-6408-4181}\inst{\ref{aff10},\ref{aff11}}
\and T.~Saifollahi\orcid{0000-0002-9554-7660}\inst{\ref{aff1}}
\and K.~Voggel\orcid{0000-0001-6215-0950}\inst{\ref{aff1}}
\and A.~Lan\c{c}on\orcid{0000-0002-7214-8296}\inst{\ref{aff1}}
\and F.~R.~Marleau\orcid{0000-0002-1442-2947}\inst{\ref{aff12}}
\and E.~Sola\orcid{0000-0002-2814-3578}\inst{\ref{aff13}}
\and L.~K.~Hunt\orcid{0000-0001-9162-2371}\inst{\ref{aff14}}
\and J.~Junais\orcid{0000-0002-7016-4532}\inst{\ref{aff15},\ref{aff16}}
\and D.~Carollo\orcid{0000-0002-0005-5787}\inst{\ref{aff9}}
\and P.~M.~Sanchez-Alarcon\orcid{0000-0002-6278-9233}\inst{\ref{aff15},\ref{aff16}}
\and M.~Baes\orcid{0000-0002-3930-2757}\inst{\ref{aff17}}
\and F.~Buitrago\orcid{0000-0002-2861-9812}\inst{\ref{aff18},\ref{aff19},\ref{aff20}}
\and Michele~Cantiello\orcid{0000-0003-2072-384X}\inst{\ref{aff21}}
\and J.-C.~Cuillandre\orcid{0000-0002-3263-8645}\inst{\ref{aff22}}
\and H.~Dom\'inguez~S\'anchez\orcid{0000-0002-9013-1316}\inst{\ref{aff23}}
\and A.~Ferr\'e-Mateu\orcid{0000-0002-6411-220X}\inst{\ref{aff15},\ref{aff16}}
\and A.~Franco\orcid{0000-0002-4761-366X}\inst{\ref{aff4},\ref{aff3},\ref{aff5}}
\and J.~Gracia-Carpio\inst{\ref{aff24}}
\and R.~Habas\orcid{0000-0002-4033-3841}\inst{\ref{aff21}}
\and M.~Hilker\orcid{0000-0002-2363-5522}\inst{\ref{aff25}}
\and E.~Iodice\orcid{0000-0003-4291-0005}\inst{\ref{aff26}}
\and J.~H.~Knapen\orcid{0000-0003-1643-0024}\inst{\ref{aff15},\ref{aff16}}
\and M.~N.~Le\orcid{0009-0003-0674-9813}\inst{\ref{aff15},\ref{aff16}}
\and D.~Mart{\'\i}nez-Delgado\inst{\ref{aff27}}
\and O.~M\"uller\orcid{0000-0003-4552-9808}\inst{\ref{aff28},\ref{aff13},\ref{aff29}}
\and F.~De~Paolis\orcid{0000-0001-6460-7563}\inst{\ref{aff3},\ref{aff4},\ref{aff5}}
\and P.~Papaderos\orcid{0000-0002-3733-8174}\inst{\ref{aff30}}
\and R.~Ragusa\inst{\ref{aff26}}
\and J.~Rom\'an\orcid{0000-0002-3849-3467}\inst{\ref{aff31}}
\and E.~Saremi\orcid{0000-0002-5075-1764}\inst{\ref{aff32}}
\and V.~Testa\orcid{0000-0003-1033-1340}\inst{\ref{aff33}}
\and B.~Altieri\orcid{0000-0003-3936-0284}\inst{\ref{aff34}}
\and L.~Amendola\orcid{0000-0002-0835-233X}\inst{\ref{aff35}}
\and S.~Andreon\orcid{0000-0002-2041-8784}\inst{\ref{aff36}}
\and N.~Auricchio\orcid{0000-0003-4444-8651}\inst{\ref{aff37}}
\and C.~Baccigalupi\orcid{0000-0002-8211-1630}\inst{\ref{aff38},\ref{aff9},\ref{aff39},\ref{aff40}}
\and M.~Baldi\orcid{0000-0003-4145-1943}\inst{\ref{aff41},\ref{aff37},\ref{aff42}}
\and S.~Bardelli\orcid{0000-0002-8900-0298}\inst{\ref{aff37}}
\and P.~Battaglia\orcid{0000-0002-7337-5909}\inst{\ref{aff37}}
\and A.~Biviano\orcid{0000-0002-0857-0732}\inst{\ref{aff9},\ref{aff38}}
\and E.~Branchini\orcid{0000-0002-0808-6908}\inst{\ref{aff43},\ref{aff44},\ref{aff36}}
\and M.~Brescia\orcid{0000-0001-9506-5680}\inst{\ref{aff45},\ref{aff26}}
\and S.~Camera\orcid{0000-0003-3399-3574}\inst{\ref{aff46},\ref{aff47},\ref{aff48}}
\and G.~Ca\~nas-Herrera\orcid{0000-0003-2796-2149}\inst{\ref{aff49},\ref{aff50}}
\and V.~Capobianco\orcid{0000-0002-3309-7692}\inst{\ref{aff48}}
\and C.~Carbone\orcid{0000-0003-0125-3563}\inst{\ref{aff51}}
\and J.~Carretero\orcid{0000-0002-3130-0204}\inst{\ref{aff52},\ref{aff53}}
\and S.~Casas\orcid{0000-0002-4751-5138}\inst{\ref{aff54}}
\and M.~Castellano\orcid{0000-0001-9875-8263}\inst{\ref{aff33}}
\and G.~Castignani\orcid{0000-0001-6831-0687}\inst{\ref{aff37}}
\and S.~Cavuoti\orcid{0000-0002-3787-4196}\inst{\ref{aff26},\ref{aff55}}
\and A.~Cimatti\inst{\ref{aff56}}
\and C.~Colodro-Conde\inst{\ref{aff15}}
\and G.~Congedo\orcid{0000-0003-2508-0046}\inst{\ref{aff57}}
\and C.~J.~Conselice\orcid{0000-0003-1949-7638}\inst{\ref{aff11}}
\and L.~Conversi\orcid{0000-0002-6710-8476}\inst{\ref{aff58},\ref{aff34}}
\and Y.~Copin\orcid{0000-0002-5317-7518}\inst{\ref{aff59}}
\and F.~Courbin\orcid{0000-0003-0758-6510}\inst{\ref{aff60},\ref{aff61}}
\and H.~M.~Courtois\orcid{0000-0003-0509-1776}\inst{\ref{aff62}}
\and M.~Cropper\orcid{0000-0003-4571-9468}\inst{\ref{aff63}}
\and A.~Da~Silva\orcid{0000-0002-6385-1609}\inst{\ref{aff64},\ref{aff65}}
\and H.~Degaudenzi\orcid{0000-0002-5887-6799}\inst{\ref{aff66}}
\and G.~De~Lucia\orcid{0000-0002-6220-9104}\inst{\ref{aff9}}
\and H.~Dole\orcid{0000-0002-9767-3839}\inst{\ref{aff67}}
\and F.~Dubath\orcid{0000-0002-6533-2810}\inst{\ref{aff66}}
\and C.~A.~J.~Duncan\orcid{0009-0003-3573-0791}\inst{\ref{aff57}}
\and X.~Dupac\inst{\ref{aff34}}
\and S.~Dusini\orcid{0000-0002-1128-0664}\inst{\ref{aff68}}
\and S.~Escoffier\orcid{0000-0002-2847-7498}\inst{\ref{aff69}}
\and M.~Farina\orcid{0000-0002-3089-7846}\inst{\ref{aff70}}
\and R.~Farinelli\inst{\ref{aff37}}
\and S.~Ferriol\inst{\ref{aff59}}
\and F.~Finelli\orcid{0000-0002-6694-3269}\inst{\ref{aff37},\ref{aff71}}
\and M.~Frailis\orcid{0000-0002-7400-2135}\inst{\ref{aff9}}
\and E.~Franceschi\orcid{0000-0002-0585-6591}\inst{\ref{aff37}}
\and M.~Fumana\orcid{0000-0001-6787-5950}\inst{\ref{aff51}}
\and S.~Galeotta\orcid{0000-0002-3748-5115}\inst{\ref{aff9}}
\and K.~George\orcid{0000-0002-1734-8455}\inst{\ref{aff72}}
\and B.~Gillis\orcid{0000-0002-4478-1270}\inst{\ref{aff57}}
\and C.~Giocoli\orcid{0000-0002-9590-7961}\inst{\ref{aff37},\ref{aff42}}
\and A.~Grazian\orcid{0000-0002-5688-0663}\inst{\ref{aff73}}
\and F.~Grupp\inst{\ref{aff24},\ref{aff6}}
\and L.~Guzzo\orcid{0000-0001-8264-5192}\inst{\ref{aff74},\ref{aff36},\ref{aff75}}
\and S.~V.~H.~Haugan\orcid{0000-0001-9648-7260}\inst{\ref{aff76}}
\and W.~Holmes\inst{\ref{aff77}}
\and I.~M.~Hook\orcid{0000-0002-2960-978X}\inst{\ref{aff78}}
\and F.~Hormuth\inst{\ref{aff79}}
\and A.~Hornstrup\orcid{0000-0002-3363-0936}\inst{\ref{aff80},\ref{aff81}}
\and K.~Jahnke\orcid{0000-0003-3804-2137}\inst{\ref{aff82}}
\and M.~Jhabvala\inst{\ref{aff83}}
\and B.~Joachimi\orcid{0000-0001-7494-1303}\inst{\ref{aff84}}
\and E.~Keih\"anen\orcid{0000-0003-1804-7715}\inst{\ref{aff85}}
\and S.~Kermiche\orcid{0000-0002-0302-5735}\inst{\ref{aff69}}
\and A.~Kiessling\orcid{0000-0002-2590-1273}\inst{\ref{aff77}}
\and B.~Kubik\orcid{0009-0006-5823-4880}\inst{\ref{aff59}}
\and M.~Kunz\orcid{0000-0002-3052-7394}\inst{\ref{aff86}}
\and H.~Kurki-Suonio\orcid{0000-0002-4618-3063}\inst{\ref{aff87},\ref{aff88}}
\and R.~Laureijs\inst{\ref{aff89}}
\and A.~M.~C.~Le~Brun\orcid{0000-0002-0936-4594}\inst{\ref{aff90}}
\and S.~Ligori\orcid{0000-0003-4172-4606}\inst{\ref{aff48}}
\and P.~B.~Lilje\orcid{0000-0003-4324-7794}\inst{\ref{aff76}}
\and V.~Lindholm\orcid{0000-0003-2317-5471}\inst{\ref{aff87},\ref{aff88}}
\and I.~Lloro\orcid{0000-0001-5966-1434}\inst{\ref{aff91}}
\and G.~Mainetti\orcid{0000-0003-2384-2377}\inst{\ref{aff92}}
\and D.~Maino\inst{\ref{aff74},\ref{aff51},\ref{aff75}}
\and E.~Maiorano\orcid{0000-0003-2593-4355}\inst{\ref{aff37}}
\and O.~Mansutti\orcid{0000-0001-5758-4658}\inst{\ref{aff9}}
\and O.~Marggraf\orcid{0000-0001-7242-3852}\inst{\ref{aff93}}
\and M.~Martinelli\orcid{0000-0002-6943-7732}\inst{\ref{aff33},\ref{aff94}}
\and N.~Martinet\orcid{0000-0003-2786-7790}\inst{\ref{aff95}}
\and F.~Marulli\orcid{0000-0002-8850-0303}\inst{\ref{aff96},\ref{aff37},\ref{aff42}}
\and R.~J.~Massey\orcid{0000-0002-6085-3780}\inst{\ref{aff97}}
\and E.~Medinaceli\orcid{0000-0002-4040-7783}\inst{\ref{aff37}}
\and S.~Mei\orcid{0000-0002-2849-559X}\inst{\ref{aff98},\ref{aff99}}
\and Y.~Mellier\inst{\ref{aff100},\ref{aff8}}
\and M.~Meneghetti\orcid{0000-0003-1225-7084}\inst{\ref{aff37},\ref{aff42}}
\and E.~Merlin\orcid{0000-0001-6870-8900}\inst{\ref{aff33}}
\and G.~Meylan\inst{\ref{aff28}}
\and A.~Mora\orcid{0000-0002-1922-8529}\inst{\ref{aff101}}
\and M.~Moresco\orcid{0000-0002-7616-7136}\inst{\ref{aff96},\ref{aff37}}
\and L.~Moscardini\orcid{0000-0002-3473-6716}\inst{\ref{aff96},\ref{aff37},\ref{aff42}}
\and R.~Nakajima\orcid{0009-0009-1213-7040}\inst{\ref{aff93}}
\and C.~Neissner\orcid{0000-0001-8524-4968}\inst{\ref{aff7},\ref{aff53}}
\and S.-M.~Niemi\orcid{0009-0005-0247-0086}\inst{\ref{aff49}}
\and C.~Padilla\orcid{0000-0001-7951-0166}\inst{\ref{aff7}}
\and S.~Paltani\orcid{0000-0002-8108-9179}\inst{\ref{aff66}}
\and F.~Pasian\orcid{0000-0002-4869-3227}\inst{\ref{aff9}}
\and K.~Pedersen\inst{\ref{aff102}}
\and V.~Pettorino\orcid{0000-0002-4203-9320}\inst{\ref{aff49}}
\and S.~Pires\orcid{0000-0002-0249-2104}\inst{\ref{aff22}}
\and G.~Polenta\orcid{0000-0003-4067-9196}\inst{\ref{aff103}}
\and M.~Poncet\inst{\ref{aff104}}
\and L.~A.~Popa\inst{\ref{aff105}}
\and L.~Pozzetti\orcid{0000-0001-7085-0412}\inst{\ref{aff37}}
\and F.~Raison\orcid{0000-0002-7819-6918}\inst{\ref{aff24}}
\and R.~Rebolo\orcid{0000-0003-3767-7085}\inst{\ref{aff15},\ref{aff106},\ref{aff16}}
\and A.~Renzi\orcid{0000-0001-9856-1970}\inst{\ref{aff107},\ref{aff68}}
\and J.~Rhodes\orcid{0000-0002-4485-8549}\inst{\ref{aff77}}
\and G.~Riccio\inst{\ref{aff26}}
\and M.~Roncarelli\orcid{0000-0001-9587-7822}\inst{\ref{aff37}}
\and R.~Saglia\orcid{0000-0003-0378-7032}\inst{\ref{aff6},\ref{aff24}}
\and Z.~Sakr\orcid{0000-0002-4823-3757}\inst{\ref{aff35},\ref{aff108},\ref{aff109}}
\and D.~Sapone\orcid{0000-0001-7089-4503}\inst{\ref{aff110}}
\and B.~Sartoris\orcid{0000-0003-1337-5269}\inst{\ref{aff6},\ref{aff9}}
\and P.~Schneider\orcid{0000-0001-8561-2679}\inst{\ref{aff93}}
\and T.~Schrabback\orcid{0000-0002-6987-7834}\inst{\ref{aff12}}
\and A.~Secroun\orcid{0000-0003-0505-3710}\inst{\ref{aff69}}
\and G.~Seidel\orcid{0000-0003-2907-353X}\inst{\ref{aff82}}
\and S.~Serrano\orcid{0000-0002-0211-2861}\inst{\ref{aff111},\ref{aff112},\ref{aff113}}
\and P.~Simon\inst{\ref{aff93}}
\and C.~Sirignano\orcid{0000-0002-0995-7146}\inst{\ref{aff107},\ref{aff68}}
\and G.~Sirri\orcid{0000-0003-2626-2853}\inst{\ref{aff42}}
\and L.~Stanco\orcid{0000-0002-9706-5104}\inst{\ref{aff68}}
\and J.-L.~Starck\orcid{0000-0003-2177-7794}\inst{\ref{aff22}}
\and J.~Steinwagner\orcid{0000-0001-7443-1047}\inst{\ref{aff24}}
\and P.~Tallada-Cresp\'{i}\orcid{0000-0002-1336-8328}\inst{\ref{aff52},\ref{aff53}}
\and A.~N.~Taylor\inst{\ref{aff57}}
\and H.~I.~Teplitz\orcid{0000-0002-7064-5424}\inst{\ref{aff114}}
\and I.~Tereno\orcid{0000-0002-4537-6218}\inst{\ref{aff64},\ref{aff20}}
\and N.~Tessore\orcid{0000-0002-9696-7931}\inst{\ref{aff84}}
\and S.~Toft\orcid{0000-0003-3631-7176}\inst{\ref{aff115},\ref{aff116}}
\and R.~Toledo-Moreo\orcid{0000-0002-2997-4859}\inst{\ref{aff117}}
\and F.~Torradeflot\orcid{0000-0003-1160-1517}\inst{\ref{aff53},\ref{aff52}}
\and I.~Tutusaus\orcid{0000-0002-3199-0399}\inst{\ref{aff113},\ref{aff111},\ref{aff108}}
\and L.~Valenziano\orcid{0000-0002-1170-0104}\inst{\ref{aff37},\ref{aff71}}
\and J.~Valiviita\orcid{0000-0001-6225-3693}\inst{\ref{aff87},\ref{aff88}}
\and T.~Vassallo\orcid{0000-0001-6512-6358}\inst{\ref{aff9},\ref{aff72}}
\and G.~Verdoes~Kleijn\orcid{0000-0001-5803-2580}\inst{\ref{aff89}}
\and A.~Veropalumbo\orcid{0000-0003-2387-1194}\inst{\ref{aff36},\ref{aff44},\ref{aff43}}
\and Y.~Wang\orcid{0000-0002-4749-2984}\inst{\ref{aff114}}
\and J.~Weller\orcid{0000-0002-8282-2010}\inst{\ref{aff6},\ref{aff24}}
\and G.~Zamorani\orcid{0000-0002-2318-301X}\inst{\ref{aff37}}
\and I.~A.~Zinchenko\orcid{0000-0002-2944-2449}\inst{\ref{aff118}}
\and E.~Zucca\orcid{0000-0002-5845-8132}\inst{\ref{aff37}}
\and M.~Ballardini\orcid{0000-0003-4481-3559}\inst{\ref{aff119},\ref{aff120},\ref{aff37}}
\and M.~Bolzonella\orcid{0000-0003-3278-4607}\inst{\ref{aff37}}
\and E.~Bozzo\orcid{0000-0002-8201-1525}\inst{\ref{aff66}}
\and C.~Burigana\orcid{0000-0002-3005-5796}\inst{\ref{aff121},\ref{aff71}}
\and R.~Cabanac\orcid{0000-0001-6679-2600}\inst{\ref{aff108}}
\and A.~Cappi\inst{\ref{aff37},\ref{aff122}}
\and D.~Di~Ferdinando\inst{\ref{aff42}}
\and J.~A.~Escartin~Vigo\inst{\ref{aff24}}
\and L.~Gabarra\orcid{0000-0002-8486-8856}\inst{\ref{aff123}}
\and M.~Huertas-Company\orcid{0000-0002-1416-8483}\inst{\ref{aff15},\ref{aff124},\ref{aff125},\ref{aff126}}
\and J.~Mart\'{i}n-Fleitas\orcid{0000-0002-8594-569X}\inst{\ref{aff127}}
\and S.~Matthew\orcid{0000-0001-8448-1697}\inst{\ref{aff57}}
\and N.~Mauri\orcid{0000-0001-8196-1548}\inst{\ref{aff56},\ref{aff42}}
\and R.~B.~Metcalf\orcid{0000-0003-3167-2574}\inst{\ref{aff96},\ref{aff37}}
\and A.~Pezzotta\orcid{0000-0003-0726-2268}\inst{\ref{aff36}}
\and M.~P\"ontinen\orcid{0000-0001-5442-2530}\inst{\ref{aff87}}
\and C.~Porciani\orcid{0000-0002-7797-2508}\inst{\ref{aff93}}
\and I.~Risso\orcid{0000-0003-2525-7761}\inst{\ref{aff36},\ref{aff44}}
\and V.~Scottez\orcid{0009-0008-3864-940X}\inst{\ref{aff100},\ref{aff128}}
\and M.~Sereno\orcid{0000-0003-0302-0325}\inst{\ref{aff37},\ref{aff42}}
\and M.~Tenti\orcid{0000-0002-4254-5901}\inst{\ref{aff42}}
\and M.~Viel\orcid{0000-0002-2642-5707}\inst{\ref{aff38},\ref{aff9},\ref{aff40},\ref{aff39},\ref{aff129}}
\and M.~Wiesmann\orcid{0009-0000-8199-5860}\inst{\ref{aff76}}
\and Y.~Akrami\orcid{0000-0002-2407-7956}\inst{\ref{aff130},\ref{aff131}}
\and I.~T.~Andika\orcid{0000-0001-6102-9526}\inst{\ref{aff132},\ref{aff133}}
\and S.~Anselmi\orcid{0000-0002-3579-9583}\inst{\ref{aff68},\ref{aff107},\ref{aff134}}
\and M.~Archidiacono\orcid{0000-0003-4952-9012}\inst{\ref{aff74},\ref{aff75}}
\and F.~Atrio-Barandela\orcid{0000-0002-2130-2513}\inst{\ref{aff135}}
\and D.~Bertacca\orcid{0000-0002-2490-7139}\inst{\ref{aff107},\ref{aff73},\ref{aff68}}
\and M.~Bethermin\orcid{0000-0002-3915-2015}\inst{\ref{aff1}}
\and A.~Blanchard\orcid{0000-0001-8555-9003}\inst{\ref{aff108}}
\and L.~Blot\orcid{0000-0002-9622-7167}\inst{\ref{aff136},\ref{aff90}}
\and M.~Bonici\orcid{0000-0002-8430-126X}\inst{\ref{aff137},\ref{aff51}}
\and S.~Borgani\orcid{0000-0001-6151-6439}\inst{\ref{aff138},\ref{aff38},\ref{aff9},\ref{aff39},\ref{aff129}}
\and M.~L.~Brown\orcid{0000-0002-0370-8077}\inst{\ref{aff11}}
\and S.~Bruton\orcid{0000-0002-6503-5218}\inst{\ref{aff139}}
\and A.~Calabro\orcid{0000-0003-2536-1614}\inst{\ref{aff33}}
\and B.~Camacho~Quevedo\orcid{0000-0002-8789-4232}\inst{\ref{aff38},\ref{aff40},\ref{aff9}}
\and F.~Caro\inst{\ref{aff33}}
\and C.~S.~Carvalho\inst{\ref{aff20}}
\and T.~Castro\orcid{0000-0002-6292-3228}\inst{\ref{aff9},\ref{aff39},\ref{aff38},\ref{aff129}}
\and F.~Cogato\orcid{0000-0003-4632-6113}\inst{\ref{aff96},\ref{aff37}}
\and S.~Conseil\orcid{0000-0002-3657-4191}\inst{\ref{aff59}}
\and A.~R.~Cooray\orcid{0000-0002-3892-0190}\inst{\ref{aff140}}
\and O.~Cucciati\orcid{0000-0002-9336-7551}\inst{\ref{aff37}}
\and S.~Davini\orcid{0000-0003-3269-1718}\inst{\ref{aff44}}
\and G.~Desprez\orcid{0000-0001-8325-1742}\inst{\ref{aff89}}
\and A.~D\'iaz-S\'anchez\orcid{0000-0003-0748-4768}\inst{\ref{aff141}}
\and J.~J.~Diaz\orcid{0000-0003-2101-1078}\inst{\ref{aff15}}
\and S.~Di~Domizio\orcid{0000-0003-2863-5895}\inst{\ref{aff43},\ref{aff44}}
\and J.~M.~Diego\orcid{0000-0001-9065-3926}\inst{\ref{aff23}}
\and M.~Y.~Elkhashab\orcid{0000-0001-9306-2603}\inst{\ref{aff9},\ref{aff39},\ref{aff138},\ref{aff38}}
\and A.~Enia\orcid{0000-0002-0200-2857}\inst{\ref{aff37},\ref{aff41}}
\and Y.~Fang\orcid{0000-0002-0334-6950}\inst{\ref{aff6}}
\and A.~G.~Ferrari\orcid{0009-0005-5266-4110}\inst{\ref{aff42}}
\and A.~Finoguenov\orcid{0000-0002-4606-5403}\inst{\ref{aff87}}
\and K.~Ganga\orcid{0000-0001-8159-8208}\inst{\ref{aff98}}
\and J.~Garc\'ia-Bellido\orcid{0000-0002-9370-8360}\inst{\ref{aff130}}
\and T.~Gasparetto\orcid{0000-0002-7913-4866}\inst{\ref{aff33}}
\and V.~Gautard\inst{\ref{aff142}}
\and E.~Gaztanaga\orcid{0000-0001-9632-0815}\inst{\ref{aff113},\ref{aff111},\ref{aff143}}
\and F.~Giacomini\orcid{0000-0002-3129-2814}\inst{\ref{aff42}}
\and F.~Gianotti\orcid{0000-0003-4666-119X}\inst{\ref{aff37}}
\and G.~Gozaliasl\orcid{0000-0002-0236-919X}\inst{\ref{aff144},\ref{aff87}}
\and M.~Guidi\orcid{0000-0001-9408-1101}\inst{\ref{aff41},\ref{aff37}}
\and C.~M.~Gutierrez\orcid{0000-0001-7854-783X}\inst{\ref{aff145}}
\and A.~Hall\orcid{0000-0002-3139-8651}\inst{\ref{aff57}}
\and H.~Hildebrandt\orcid{0000-0002-9814-3338}\inst{\ref{aff146}}
\and J.~Hjorth\orcid{0000-0002-4571-2306}\inst{\ref{aff102}}
\and J.~J.~E.~Kajava\orcid{0000-0002-3010-8333}\inst{\ref{aff147},\ref{aff148}}
\and Y.~Kang\orcid{0009-0000-8588-7250}\inst{\ref{aff66}}
\and V.~Kansal\orcid{0000-0002-4008-6078}\inst{\ref{aff149},\ref{aff150}}
\and D.~Karagiannis\orcid{0000-0002-4927-0816}\inst{\ref{aff119},\ref{aff151}}
\and K.~Kiiveri\inst{\ref{aff85}}
\and J.~Kim\orcid{0000-0003-2776-2761}\inst{\ref{aff123}}
\and C.~C.~Kirkpatrick\inst{\ref{aff85}}
\and S.~Kruk\orcid{0000-0001-8010-8879}\inst{\ref{aff34}}
\and J.~Le~Graet\orcid{0000-0001-6523-7971}\inst{\ref{aff69}}
\and L.~Legrand\orcid{0000-0003-0610-5252}\inst{\ref{aff152},\ref{aff153}}
\and M.~Lembo\orcid{0000-0002-5271-5070}\inst{\ref{aff8},\ref{aff119},\ref{aff120}}
\and F.~Lepori\orcid{0009-0000-5061-7138}\inst{\ref{aff154}}
\and G.~Leroy\orcid{0009-0004-2523-4425}\inst{\ref{aff155},\ref{aff97}}
\and G.~F.~Lesci\orcid{0000-0002-4607-2830}\inst{\ref{aff96},\ref{aff37}}
\and J.~Lesgourgues\orcid{0000-0001-7627-353X}\inst{\ref{aff54}}
\and L.~Leuzzi\orcid{0009-0006-4479-7017}\inst{\ref{aff37}}
\and T.~I.~Liaudat\orcid{0000-0002-9104-314X}\inst{\ref{aff156}}
\and A.~Loureiro\orcid{0000-0002-4371-0876}\inst{\ref{aff157},\ref{aff158}}
\and J.~Macias-Perez\orcid{0000-0002-5385-2763}\inst{\ref{aff159}}
\and G.~Maggio\orcid{0000-0003-4020-4836}\inst{\ref{aff9}}
\and M.~Magliocchetti\orcid{0000-0001-9158-4838}\inst{\ref{aff70}}
\and F.~Mannucci\orcid{0000-0002-4803-2381}\inst{\ref{aff14}}
\and R.~Maoli\orcid{0000-0002-6065-3025}\inst{\ref{aff160},\ref{aff33}}
\and C.~J.~A.~P.~Martins\orcid{0000-0002-4886-9261}\inst{\ref{aff161},\ref{aff30}}
\and L.~Maurin\orcid{0000-0002-8406-0857}\inst{\ref{aff67}}
\and M.~Miluzio\inst{\ref{aff34},\ref{aff162}}
\and P.~Monaco\orcid{0000-0003-2083-7564}\inst{\ref{aff138},\ref{aff9},\ref{aff39},\ref{aff38},\ref{aff129}}
\and C.~Moretti\orcid{0000-0003-3314-8936}\inst{\ref{aff9},\ref{aff38},\ref{aff39},\ref{aff40}}
\and G.~Morgante\inst{\ref{aff37}}
\and K.~Naidoo\orcid{0000-0002-9182-1802}\inst{\ref{aff143},\ref{aff84}}
\and A.~Navarro-Alsina\orcid{0000-0002-3173-2592}\inst{\ref{aff93}}
\and S.~Nesseris\orcid{0000-0002-0567-0324}\inst{\ref{aff130}}
\and D.~Paoletti\orcid{0000-0003-4761-6147}\inst{\ref{aff37},\ref{aff71}}
\and F.~Passalacqua\orcid{0000-0002-8606-4093}\inst{\ref{aff107},\ref{aff68}}
\and K.~Paterson\orcid{0000-0001-8340-3486}\inst{\ref{aff82}}
\and L.~Patrizii\inst{\ref{aff42}}
\and A.~Pisani\orcid{0000-0002-6146-4437}\inst{\ref{aff69}}
\and D.~Potter\orcid{0000-0002-0757-5195}\inst{\ref{aff154}}
\and S.~Quai\orcid{0000-0002-0449-8163}\inst{\ref{aff96},\ref{aff37}}
\and M.~Radovich\orcid{0000-0002-3585-866X}\inst{\ref{aff73}}
\and G.~Rodighiero\orcid{0000-0002-9415-2296}\inst{\ref{aff107},\ref{aff73}}
\and S.~Sacquegna\orcid{0000-0002-8433-6630}\inst{\ref{aff21},\ref{aff3},\ref{aff4}}
\and M.~Sahl\'en\orcid{0000-0003-0973-4804}\inst{\ref{aff163}}
\and D.~B.~Sanders\orcid{0000-0002-1233-9998}\inst{\ref{aff164}}
\and E.~Sarpa\orcid{0000-0002-1256-655X}\inst{\ref{aff40},\ref{aff129},\ref{aff39}}
\and A.~Schneider\orcid{0000-0001-7055-8104}\inst{\ref{aff154}}
\and D.~Sciotti\orcid{0009-0008-4519-2620}\inst{\ref{aff33},\ref{aff94}}
\and E.~Sellentin\inst{\ref{aff165},\ref{aff50}}
\and L.~C.~Smith\orcid{0000-0002-3259-2771}\inst{\ref{aff13}}
\and J.~G.~Sorce\orcid{0000-0002-2307-2432}\inst{\ref{aff166},\ref{aff67}}
\and K.~Tanidis\orcid{0000-0001-9843-5130}\inst{\ref{aff123}}
\and C.~Tao\orcid{0000-0001-7961-8177}\inst{\ref{aff69}}
\and G.~Testera\inst{\ref{aff44}}
\and R.~Teyssier\orcid{0000-0001-7689-0933}\inst{\ref{aff167}}
\and S.~Tosi\orcid{0000-0002-7275-9193}\inst{\ref{aff43},\ref{aff44},\ref{aff36}}
\and A.~Troja\orcid{0000-0003-0239-4595}\inst{\ref{aff107},\ref{aff68}}
\and M.~Tucci\inst{\ref{aff66}}
\and C.~Valieri\inst{\ref{aff42}}
\and D.~Vergani\orcid{0000-0003-0898-2216}\inst{\ref{aff37}}
\and G.~Verza\orcid{0000-0002-1886-8348}\inst{\ref{aff168}}
\and P.~Vielzeuf\orcid{0000-0003-2035-9339}\inst{\ref{aff69}}
\and N.~A.~Walton\orcid{0000-0003-3983-8778}\inst{\ref{aff13}}}
                                                                                   
\institute{Universit\'e de Strasbourg, CNRS, Observatoire astronomique de Strasbourg, UMR 7550, 67000 Strasbourg, France\label{aff1}
\and
Space physics and astronomy research unit, University of Oulu, Pentti Kaiteran katu 1, FI-90014 Oulu, Finland\label{aff2}
\and
Department of Mathematics and Physics E. De Giorgi, University of Salento, Via per Arnesano, CP-I93, 73100, Lecce, Italy\label{aff3}
\and
INFN, Sezione di Lecce, Via per Arnesano, CP-193, 73100, Lecce, Italy\label{aff4}
\and
INAF-Sezione di Lecce, c/o Dipartimento Matematica e Fisica, Via per Arnesano, 73100, Lecce, Italy\label{aff5}
\and
Universit\"ats-Sternwarte M\"unchen, Fakult\"at f\"ur Physik, Ludwig-Maximilians-Universit\"at M\"unchen, Scheinerstrasse 1, 81679 M\"unchen, Germany\label{aff6}
\and
Institut de F\'{i}sica d'Altes Energies (IFAE), The Barcelona Institute of Science and Technology, Campus UAB, 08193 Bellaterra (Barcelona), Spain\label{aff7}
\and
Institut d'Astrophysique de Paris, UMR 7095, CNRS, and Sorbonne Universit\'e, 98 bis boulevard Arago, 75014 Paris, France\label{aff8}
\and
INAF-Osservatorio Astronomico di Trieste, Via G. B. Tiepolo 11, 34143 Trieste, Italy\label{aff9}
\and
David A. Dunlap Department of Astronomy \& Astrophysics, University of Toronto, 50 St George Street, Toronto, Ontario M5S 3H4, Canada\label{aff10}
\and
Jodrell Bank Centre for Astrophysics, Department of Physics and Astronomy, University of Manchester, Oxford Road, Manchester M13 9PL, UK\label{aff11}
\and
Universit\"at Innsbruck, Institut f\"ur Astro- und Teilchenphysik, Technikerstr. 25/8, 6020 Innsbruck, Austria\label{aff12}
\and
Institute of Astronomy, University of Cambridge, Madingley Road, Cambridge CB3 0HA, UK\label{aff13}
\and
INAF-Osservatorio Astrofisico di Arcetri, Largo E. Fermi 5, 50125, Firenze, Italy\label{aff14}
\and
Instituto de Astrof\'{\i}sica de Canarias, V\'{\i}a L\'actea, 38205 La Laguna, Tenerife, Spain\label{aff15}
\and
Universidad de La Laguna, Departamento de Astrof\'{\i}sica, 38206 La Laguna, Tenerife, Spain\label{aff16}
\and
Sterrenkundig Observatorium, Universiteit Gent, Krijgslaan 281 S9, 9000 Gent, Belgium\label{aff17}
\and
Departamento de F\'{i}sica Te\'{o}rica, At\'{o}mica y \'{O}ptica, Universidad de Valladolid, 47011 Valladolid, Spain\label{aff18}
\and
Laboratory for Disruptive Interdisciplinary Science (LaDIS), Universidad de Valladolid, 47011 Valladolid, Spain\label{aff19}
\and
Instituto de Astrof\'isica e Ci\^encias do Espa\c{c}o, Faculdade de Ci\^encias, Universidade de Lisboa, Tapada da Ajuda, 1349-018 Lisboa, Portugal\label{aff20}
\and
INAF - Osservatorio Astronomico d'Abruzzo, Via Maggini, 64100, Teramo, Italy\label{aff21}
\and
Universit\'e Paris-Saclay, Universit\'e Paris Cit\'e, CEA, CNRS, AIM, 91191, Gif-sur-Yvette, France\label{aff22}
\and
Instituto de F\'isica de Cantabria, Edificio Juan Jord\'a, Avenida de los Castros, 39005 Santander, Spain\label{aff23}
\and
Max Planck Institute for Extraterrestrial Physics, Giessenbachstr. 1, 85748 Garching, Germany\label{aff24}
\and
European Southern Observatory, Karl-Schwarzschild-Str.~2, 85748 Garching, Germany\label{aff25}
\and
INAF-Osservatorio Astronomico di Capodimonte, Via Moiariello 16, 80131 Napoli, Italy\label{aff26}
\and
Instituto de Astrof\'isica de Andaluc\'ia, CSIC, Glorieta de la Astronom\'\i a, 18080, Granada, Spain\label{aff27}
\and
Institute of Physics, Laboratory of Astrophysics, Ecole Polytechnique F\'ed\'erale de Lausanne (EPFL), Observatoire de Sauverny, 1290 Versoix, Switzerland\label{aff28}
\and
Visiting Fellow, Clare Hall, University of Cambridge, Cambridge, UK\label{aff29}
\and
Instituto de Astrof\'isica e Ci\^encias do Espa\c{c}o, Universidade do Porto, CAUP, Rua das Estrelas, PT4150-762 Porto, Portugal\label{aff30}
\and
Departamento de F{\'\i}sica de la Tierra y Astrof{\'\i}sica, Universidad Complutense de Madrid, Plaza de las Ciencias 2, E-28040 Madrid, Spain\label{aff31}
\and
School of Physics \& Astronomy, University of Southampton, Highfield Campus, Southampton SO17 1BJ, UK\label{aff32}
\and
INAF-Osservatorio Astronomico di Roma, Via Frascati 33, 00078 Monteporzio Catone, Italy\label{aff33}
\and
ESAC/ESA, Camino Bajo del Castillo, s/n., Urb. Villafranca del Castillo, 28692 Villanueva de la Ca\~nada, Madrid, Spain\label{aff34}
\and
Institut f\"ur Theoretische Physik, University of Heidelberg, Philosophenweg 16, 69120 Heidelberg, Germany\label{aff35}
\and
INAF-Osservatorio Astronomico di Brera, Via Brera 28, 20122 Milano, Italy\label{aff36}
\and
INAF-Osservatorio di Astrofisica e Scienza dello Spazio di Bologna, Via Piero Gobetti 93/3, 40129 Bologna, Italy\label{aff37}
\and
IFPU, Institute for Fundamental Physics of the Universe, via Beirut 2, 34151 Trieste, Italy\label{aff38}
\and
INFN, Sezione di Trieste, Via Valerio 2, 34127 Trieste TS, Italy\label{aff39}
\and
SISSA, International School for Advanced Studies, Via Bonomea 265, 34136 Trieste TS, Italy\label{aff40}
\and
Dipartimento di Fisica e Astronomia, Universit\`a di Bologna, Via Gobetti 93/2, 40129 Bologna, Italy\label{aff41}
\and
INFN-Sezione di Bologna, Viale Berti Pichat 6/2, 40127 Bologna, Italy\label{aff42}
\and
Dipartimento di Fisica, Universit\`a di Genova, Via Dodecaneso 33, 16146, Genova, Italy\label{aff43}
\and
INFN-Sezione di Genova, Via Dodecaneso 33, 16146, Genova, Italy\label{aff44}
\and
Department of Physics "E. Pancini", University Federico II, Via Cinthia 6, 80126, Napoli, Italy\label{aff45}
\and
Dipartimento di Fisica, Universit\`a degli Studi di Torino, Via P. Giuria 1, 10125 Torino, Italy\label{aff46}
\and
INFN-Sezione di Torino, Via P. Giuria 1, 10125 Torino, Italy\label{aff47}
\and
INAF-Osservatorio Astrofisico di Torino, Via Osservatorio 20, 10025 Pino Torinese (TO), Italy\label{aff48}
\and
European Space Agency/ESTEC, Keplerlaan 1, 2201 AZ Noordwijk, The Netherlands\label{aff49}
\and
Leiden Observatory, Leiden University, Einsteinweg 55, 2333 CC Leiden, The Netherlands\label{aff50}
\and
INAF-IASF Milano, Via Alfonso Corti 12, 20133 Milano, Italy\label{aff51}
\and
Centro de Investigaciones Energ\'eticas, Medioambientales y Tecnol\'ogicas (CIEMAT), Avenida Complutense 40, 28040 Madrid, Spain\label{aff52}
\and
Port d'Informaci\'{o} Cient\'{i}fica, Campus UAB, C. Albareda s/n, 08193 Bellaterra (Barcelona), Spain\label{aff53}
\and
Institute for Theoretical Particle Physics and Cosmology (TTK), RWTH Aachen University, 52056 Aachen, Germany\label{aff54}
\and
INFN section of Naples, Via Cinthia 6, 80126, Napoli, Italy\label{aff55}
\and
Dipartimento di Fisica e Astronomia "Augusto Righi" - Alma Mater Studiorum Universit\`a di Bologna, Viale Berti Pichat 6/2, 40127 Bologna, Italy\label{aff56}
\and
Institute for Astronomy, University of Edinburgh, Royal Observatory, Blackford Hill, Edinburgh EH9 3HJ, UK\label{aff57}
\and
European Space Agency/ESRIN, Largo Galileo Galilei 1, 00044 Frascati, Roma, Italy\label{aff58}
\and
Universit\'e Claude Bernard Lyon 1, CNRS/IN2P3, IP2I Lyon, UMR 5822, Villeurbanne, F-69100, France\label{aff59}
\and
Institut de Ci\`{e}ncies del Cosmos (ICCUB), Universitat de Barcelona (IEEC-UB), Mart\'{i} i Franqu\`{e}s 1, 08028 Barcelona, Spain\label{aff60}
\and
Instituci\'o Catalana de Recerca i Estudis Avan\c{c}ats (ICREA), Passeig de Llu\'{\i}s Companys 23, 08010 Barcelona, Spain\label{aff61}
\and
UCB Lyon 1, CNRS/IN2P3, IUF, IP2I Lyon, 4 rue Enrico Fermi, 69622 Villeurbanne, France\label{aff62}
\and
Mullard Space Science Laboratory, University College London, Holmbury St Mary, Dorking, Surrey RH5 6NT, UK\label{aff63}
\and
Departamento de F\'isica, Faculdade de Ci\^encias, Universidade de Lisboa, Edif\'icio C8, Campo Grande, PT1749-016 Lisboa, Portugal\label{aff64}
\and
Instituto de Astrof\'isica e Ci\^encias do Espa\c{c}o, Faculdade de Ci\^encias, Universidade de Lisboa, Campo Grande, 1749-016 Lisboa, Portugal\label{aff65}
\and
Department of Astronomy, University of Geneva, ch. d'Ecogia 16, 1290 Versoix, Switzerland\label{aff66}
\and
Universit\'e Paris-Saclay, CNRS, Institut d'astrophysique spatiale, 91405, Orsay, France\label{aff67}
\and
INFN-Padova, Via Marzolo 8, 35131 Padova, Italy\label{aff68}
\and
Aix-Marseille Universit\'e, CNRS/IN2P3, CPPM, Marseille, France\label{aff69}
\and
INAF-Istituto di Astrofisica e Planetologia Spaziali, via del Fosso del Cavaliere, 100, 00100 Roma, Italy\label{aff70}
\and
INFN-Bologna, Via Irnerio 46, 40126 Bologna, Italy\label{aff71}
\and
University Observatory, LMU Faculty of Physics, Scheinerstrasse 1, 81679 Munich, Germany\label{aff72}
\and
INAF-Osservatorio Astronomico di Padova, Via dell'Osservatorio 5, 35122 Padova, Italy\label{aff73}
\and
Dipartimento di Fisica "Aldo Pontremoli", Universit\`a degli Studi di Milano, Via Celoria 16, 20133 Milano, Italy\label{aff74}
\and
INFN-Sezione di Milano, Via Celoria 16, 20133 Milano, Italy\label{aff75}
\and
Institute of Theoretical Astrophysics, University of Oslo, P.O. Box 1029 Blindern, 0315 Oslo, Norway\label{aff76}
\and
Jet Propulsion Laboratory, California Institute of Technology, 4800 Oak Grove Drive, Pasadena, CA, 91109, USA\label{aff77}
\and
Department of Physics, Lancaster University, Lancaster, LA1 4YB, UK\label{aff78}
\and
Felix Hormuth Engineering, Goethestr. 17, 69181 Leimen, Germany\label{aff79}
\and
Technical University of Denmark, Elektrovej 327, 2800 Kgs. Lyngby, Denmark\label{aff80}
\and
Cosmic Dawn Center (DAWN), Denmark\label{aff81}
\and
Max-Planck-Institut f\"ur Astronomie, K\"onigstuhl 17, 69117 Heidelberg, Germany\label{aff82}
\and
NASA Goddard Space Flight Center, Greenbelt, MD 20771, USA\label{aff83}
\and
Department of Physics and Astronomy, University College London, Gower Street, London WC1E 6BT, UK\label{aff84}
\and
Department of Physics and Helsinki Institute of Physics, Gustaf H\"allstr\"omin katu 2, University of Helsinki, 00014 Helsinki, Finland\label{aff85}
\and
Universit\'e de Gen\`eve, D\'epartement de Physique Th\'eorique and Centre for Astroparticle Physics, 24 quai Ernest-Ansermet, CH-1211 Gen\`eve 4, Switzerland\label{aff86}
\and
Department of Physics, P.O. Box 64, University of Helsinki, 00014 Helsinki, Finland\label{aff87}
\and
Helsinki Institute of Physics, Gustaf H{\"a}llstr{\"o}min katu 2, University of Helsinki, 00014 Helsinki, Finland\label{aff88}
\and
Kapteyn Astronomical Institute, University of Groningen, PO Box 800, 9700 AV Groningen, The Netherlands\label{aff89}
\and
Laboratoire d'etude de l'Univers et des phenomenes eXtremes, Observatoire de Paris, Universit\'e PSL, Sorbonne Universit\'e, CNRS, 92190 Meudon, France\label{aff90}
\and
SKAO, Jodrell Bank, Lower Withington, Macclesfield SK11 9FT, United Kingdom\label{aff91}
\and
Centre de Calcul de l'IN2P3/CNRS, 21 avenue Pierre de Coubertin 69627 Villeurbanne Cedex, France\label{aff92}
\and
Universit\"at Bonn, Argelander-Institut f\"ur Astronomie, Auf dem H\"ugel 71, 53121 Bonn, Germany\label{aff93}
\and
INFN-Sezione di Roma, Piazzale Aldo Moro, 2 - c/o Dipartimento di Fisica, Edificio G. Marconi, 00185 Roma, Italy\label{aff94}
\and
Aix-Marseille Universit\'e, CNRS, CNES, LAM, Marseille, France\label{aff95}
\and
Dipartimento di Fisica e Astronomia "Augusto Righi" - Alma Mater Studiorum Universit\`a di Bologna, via Piero Gobetti 93/2, 40129 Bologna, Italy\label{aff96}
\and
Department of Physics, Institute for Computational Cosmology, Durham University, South Road, Durham, DH1 3LE, UK\label{aff97}
\and
Universit\'e Paris Cit\'e, CNRS, Astroparticule et Cosmologie, 75013 Paris, France\label{aff98}
\and
CNRS-UCB International Research Laboratory, Centre Pierre Bin\'etruy, IRL2007, CPB-IN2P3, Berkeley, USA\label{aff99}
\and
Institut d'Astrophysique de Paris, 98bis Boulevard Arago, 75014, Paris, France\label{aff100}
\and
Telespazio UK S.L. for European Space Agency (ESA), Camino bajo del Castillo, s/n, Urbanizacion Villafranca del Castillo, Villanueva de la Ca\~nada, 28692 Madrid, Spain\label{aff101}
\and
DARK, Niels Bohr Institute, University of Copenhagen, Jagtvej 155, 2200 Copenhagen, Denmark\label{aff102}
\and
Space Science Data Center, Italian Space Agency, via del Politecnico snc, 00133 Roma, Italy\label{aff103}
\and
Centre National d'Etudes Spatiales -- Centre spatial de Toulouse, 18 avenue Edouard Belin, 31401 Toulouse Cedex 9, France\label{aff104}
\and
Institute of Space Science, Str. Atomistilor, nr. 409 M\u{a}gurele, Ilfov, 077125, Romania\label{aff105}
\and
Consejo Superior de Investigaciones Cientificas, Calle Serrano 117, 28006 Madrid, Spain\label{aff106}
\and
Dipartimento di Fisica e Astronomia "G. Galilei", Universit\`a di Padova, Via Marzolo 8, 35131 Padova, Italy\label{aff107}
\and
Institut de Recherche en Astrophysique et Plan\'etologie (IRAP), Universit\'e de Toulouse, CNRS, UPS, CNES, 14 Av. Edouard Belin, 31400 Toulouse, France\label{aff108}
\and
Universit\'e St Joseph; Faculty of Sciences, Beirut, Lebanon\label{aff109}
\and
Departamento de F\'isica, FCFM, Universidad de Chile, Blanco Encalada 2008, Santiago, Chile\label{aff110}
\and
Institut d'Estudis Espacials de Catalunya (IEEC),  Edifici RDIT, Campus UPC, 08860 Castelldefels, Barcelona, Spain\label{aff111}
\and
Satlantis, University Science Park, Sede Bld 48940, Leioa-Bilbao, Spain\label{aff112}
\and
Institute of Space Sciences (ICE, CSIC), Campus UAB, Carrer de Can Magrans, s/n, 08193 Barcelona, Spain\label{aff113}
\and
Infrared Processing and Analysis Center, California Institute of Technology, Pasadena, CA 91125, USA\label{aff114}
\and
Cosmic Dawn Center (DAWN)\label{aff115}
\and
Niels Bohr Institute, University of Copenhagen, Jagtvej 128, 2200 Copenhagen, Denmark\label{aff116}
\and
Universidad Polit\'ecnica de Cartagena, Departamento de Electr\'onica y Tecnolog\'ia de Computadoras,  Plaza del Hospital 1, 30202 Cartagena, Spain\label{aff117}
\and
Astronomisches Rechen-Institut, Zentrum f\"ur Astronomie der Universit\"at Heidelberg, M\"onchhofstr. 12-14, 69120 Heidelberg, Germany\label{aff118}
\and
Dipartimento di Fisica e Scienze della Terra, Universit\`a degli Studi di Ferrara, Via Giuseppe Saragat 1, 44122 Ferrara, Italy\label{aff119}
\and
Istituto Nazionale di Fisica Nucleare, Sezione di Ferrara, Via Giuseppe Saragat 1, 44122 Ferrara, Italy\label{aff120}
\and
INAF, Istituto di Radioastronomia, Via Piero Gobetti 101, 40129 Bologna, Italy\label{aff121}
\and
Universit\'e C\^{o}te d'Azur, Observatoire de la C\^{o}te d'Azur, CNRS, Laboratoire Lagrange, Bd de l'Observatoire, CS 34229, 06304 Nice cedex 4, France\label{aff122}
\and
Department of Physics, Oxford University, Keble Road, Oxford OX1 3RH, UK\label{aff123}
\and
Instituto de Astrof\'isica de Canarias (IAC); Departamento de Astrof\'isica, Universidad de La Laguna (ULL), 38200, La Laguna, Tenerife, Spain\label{aff124}
\and
Universit\'e PSL, Observatoire de Paris, Sorbonne Universit\'e, CNRS, LERMA, 75014, Paris, France\label{aff125}
\and
Universit\'e Paris-Cit\'e, 5 Rue Thomas Mann, 75013, Paris, France\label{aff126}
\and
Aurora Technology for European Space Agency (ESA), Camino bajo del Castillo, s/n, Urbanizacion Villafranca del Castillo, Villanueva de la Ca\~nada, 28692 Madrid, Spain\label{aff127}
\and
ICL, Junia, Universit\'e Catholique de Lille, LITL, 59000 Lille, France\label{aff128}
\and
ICSC - Centro Nazionale di Ricerca in High Performance Computing, Big Data e Quantum Computing, Via Magnanelli 2, Bologna, Italy\label{aff129}
\and
Instituto de F\'isica Te\'orica UAM-CSIC, Campus de Cantoblanco, 28049 Madrid, Spain\label{aff130}
\and
CERCA/ISO, Department of Physics, Case Western Reserve University, 10900 Euclid Avenue, Cleveland, OH 44106, USA\label{aff131}
\and
Technical University of Munich, TUM School of Natural Sciences, Physics Department, James-Franck-Str.~1, 85748 Garching, Germany\label{aff132}
\and
Max-Planck-Institut f\"ur Astrophysik, Karl-Schwarzschild-Str.~1, 85748 Garching, Germany\label{aff133}
\and
Laboratoire Univers et Th\'eorie, Observatoire de Paris, Universit\'e PSL, Universit\'e Paris Cit\'e, CNRS, 92190 Meudon, France\label{aff134}
\and
Departamento de F{\'\i}sica Fundamental. Universidad de Salamanca. Plaza de la Merced s/n. 37008 Salamanca, Spain\label{aff135}
\and
Center for Data-Driven Discovery, Kavli IPMU (WPI), UTIAS, The University of Tokyo, Kashiwa, Chiba 277-8583, Japan\label{aff136}
\and
Waterloo Centre for Astrophysics, University of Waterloo, Waterloo, Ontario N2L 3G1, Canada\label{aff137}
\and
Dipartimento di Fisica - Sezione di Astronomia, Universit\`a di Trieste, Via Tiepolo 11, 34131 Trieste, Italy\label{aff138}
\and
California Institute of Technology, 1200 E California Blvd, Pasadena, CA 91125, USA\label{aff139}
\and
Department of Physics \& Astronomy, University of California Irvine, Irvine CA 92697, USA\label{aff140}
\and
Departamento F\'isica Aplicada, Universidad Polit\'ecnica de Cartagena, Campus Muralla del Mar, 30202 Cartagena, Murcia, Spain\label{aff141}
\and
CEA Saclay, DFR/IRFU, Service d'Astrophysique, Bat. 709, 91191 Gif-sur-Yvette, France\label{aff142}
\and
Institute of Cosmology and Gravitation, University of Portsmouth, Portsmouth PO1 3FX, UK\label{aff143}
\and
Department of Computer Science, Aalto University, PO Box 15400, Espoo, FI-00 076, Finland\label{aff144}
\and
Instituto de Astrof\'\i sica de Canarias, c/ Via Lactea s/n, La Laguna 38200, Spain. Departamento de Astrof\'\i sica de la Universidad de La Laguna, Avda. Francisco Sanchez, La Laguna, 38200, Spain\label{aff145}
\and
Ruhr University Bochum, Faculty of Physics and Astronomy, Astronomical Institute (AIRUB), German Centre for Cosmological Lensing (GCCL), 44780 Bochum, Germany\label{aff146}
\and
Department of Physics and Astronomy, Vesilinnantie 5, University of Turku, 20014 Turku, Finland\label{aff147}
\and
Serco for European Space Agency (ESA), Camino bajo del Castillo, s/n, Urbanizacion Villafranca del Castillo, Villanueva de la Ca\~nada, 28692 Madrid, Spain\label{aff148}
\and
ARC Centre of Excellence for Dark Matter Particle Physics, Melbourne, Australia\label{aff149}
\and
Centre for Astrophysics \& Supercomputing, Swinburne University of Technology,  Hawthorn, Victoria 3122, Australia\label{aff150}
\and
Department of Physics and Astronomy, University of the Western Cape, Bellville, Cape Town, 7535, South Africa\label{aff151}
\and
DAMTP, Centre for Mathematical Sciences, Wilberforce Road, Cambridge CB3 0WA, UK\label{aff152}
\and
Kavli Institute for Cosmology Cambridge, Madingley Road, Cambridge, CB3 0HA, UK\label{aff153}
\and
Department of Astrophysics, University of Zurich, Winterthurerstrasse 190, 8057 Zurich, Switzerland\label{aff154}
\and
Department of Physics, Centre for Extragalactic Astronomy, Durham University, South Road, Durham, DH1 3LE, UK\label{aff155}
\and
IRFU, CEA, Universit\'e Paris-Saclay 91191 Gif-sur-Yvette Cedex, France\label{aff156}
\and
Oskar Klein Centre for Cosmoparticle Physics, Department of Physics, Stockholm University, Stockholm, SE-106 91, Sweden\label{aff157}
\and
Astrophysics Group, Blackett Laboratory, Imperial College London, London SW7 2AZ, UK\label{aff158}
\and
Univ. Grenoble Alpes, CNRS, Grenoble INP, LPSC-IN2P3, 53, Avenue des Martyrs, 38000, Grenoble, France\label{aff159}
\and
Dipartimento di Fisica, Sapienza Universit\`a di Roma, Piazzale Aldo Moro 2, 00185 Roma, Italy\label{aff160}
\and
Centro de Astrof\'{\i}sica da Universidade do Porto, Rua das Estrelas, 4150-762 Porto, Portugal\label{aff161}
\and
HE Space for European Space Agency (ESA), Camino bajo del Castillo, s/n, Urbanizacion Villafranca del Castillo, Villanueva de la Ca\~nada, 28692 Madrid, Spain\label{aff162}
\and
Theoretical astrophysics, Department of Physics and Astronomy, Uppsala University, Box 516, 751 37 Uppsala, Sweden\label{aff163}
\and
Institute for Astronomy, University of Hawaii, 2680 Woodlawn Drive, Honolulu, HI 96822, USA\label{aff164}
\and
Mathematical Institute, University of Leiden, Einsteinweg 55, 2333 CA Leiden, The Netherlands\label{aff165}
\and
Univ. Lille, CNRS, Centrale Lille, UMR 9189 CRIStAL, 59000 Lille, France\label{aff166}
\and
Department of Astrophysical Sciences, Peyton Hall, Princeton University, Princeton, NJ 08544, USA\label{aff167}
\and
Center for Computational Astrophysics, Flatiron Institute, 162 5th Avenue, 10010, New York, NY, USA\label{aff168}}               

 \abstract{
Local Universe dwarf galaxies can serve as both cosmological and mass assembly probes. Deep surveys have enabled the study of these objects down to the low surface brightness (LSB) regime. In this paper, we estimate \Euclid's dwarf detection capabilities as well as limits of its MERge processing function (MER pipeline), which is responsible for producing the stacked mosaics and final catalogues. To do this, we injected mock dwarf galaxies in a real Euclid Wide Survey (EWS) field in the VIS band and compared the input catalogue to the final MER catalogue. The mock dwarf galaxies were generated using simple Sérsic models with structural parameters (including size and surface brightness) drawn from observed dwarf galaxy catalogues. These simulations represent an idealised case in the sense they do not account for additional factors such as ellipticity, morphology, or crowding. To characterise the detected dwarfs, we used the mean surface brightness inside the effective radius ${\rm SB}_{\rm e}$ (in mag\,arcsec$^{-2}$). The final MER catalogues achieve a completenesses of $91\,\%$ for ${\rm SB}_{\rm e}\in[21,24]$ and $54\,\%$ for ${\rm SB}_{\rm e}\in[24,28]$. These numbers do not take into account possible contaminants, including confusion with background galaxies at the location of the dwarfs. After taking those effects into account, they respectively became $86\,\%$ and $38\,\%$. The MER pipeline performs a final local background subtraction with a small mesh size, leading to a flux loss for galaxies with $R_{\rm e}>10\arcsec$. By using the final MER mosaics and reinjecting this local background, we obtained an image in which we recover reliable photometric properties for objects under the arcminute scale. This background-reinjected product is thus suitable for the study of Local Universe dwarf galaxies. \Euclid's data reduction pipeline serves as a test bed for other deep surveys, particularly regarding background subtraction methods, a key issue in LSB science.

}

    \keywords{Galaxies: dwarf -- Techniques: image processing -- Catalogues}

   \titlerunning{Using mock LSB dwarf galaxies to probe Euclid Wide Survey detection capabilities}
   \authorrunning{Euclid Collaboration: M. Urbano et al.}
   
  \maketitle
  
\section{\label{sc:Intro}Introduction}

The dark energy, cold dark matter standard model suggests that galaxies originate from the first small, low-mass haloes that were formed as a result of primordial variations in the density of cold dark matter \citep[e.g.][]{1986ApJ...303...39D}. Today, a diverse range of galaxies can be observed in the Local Universe varying in size, mass, and morphology. In particular, galaxies at the low-mass end, known as dwarf galaxies, can exhibit a low surface brightness (LSB), and they come in a variety of morphologies. Local Universe dwarf galaxies can also be used as cosmological probes, whose number and distribution around larger galaxies constrain dark matter models in simulations \citep[e.g.][]{2000PhRvL..84.3760S,2001ApJ...556...93B,2009ApJ...696.2179K,2019ApJ...873...34N}.

The detection and identification of dwarf galaxies have presented significant challenges. Over the past decades, wide surveys such as the Sloan Digital Sky Survey (SDSS, \citealt{2003AJ....126.2081A,2005ApJ...625..613A}) and Pan-STARRS \citep{2016arXiv161205560C} and deep Andromeda galaxy surveys such as the Panchromatic Hubble Andromeda Treasury (PHAT, \citealt{2012ApJS..200...18D}) and the Pan-Andromeda Archaeological Survey (PAndAS, \citealt{2009Natur.461...66M,2022ApJ...933..135D,2023ApJ...952...72D}) have extended the list of the known dwarf satellites of the Local Group to fainter luminosities and helped map their distribution. Technological advancements in terms of sensitivity in astronomical camera systems have unveiled LSB diffuse stellar features, extending the study of dwarf galaxies from the Local Group to much further in the Local Universe, where galaxies cannot be resolved into individual stars.
    
Several projects have extended the study of dwarfs to larger distances, such as the surveys conducted by the Canada-France-Hawaii telescope (including the Mass Assembly of early-Type GaLAxies with their fine Structures MATLAS: \citealt{2020MNRAS.491.1901H,Marleau2021,2021MNRAS.506.5494P,2023A&A...676A..33H}; the Canada-France Imaging Survey CFIS: \citealt{2017ApJ...848..128I}; the Next Generation Virgo Cluster Survey NGVS
: \citealt{2012ApJS..200....4F}), the Survey Telescope of the Very Large Telescope VST (including the Fornax Deep Survey, e.g. \citealt{2018A&A...620A.165V}), the Subaru telescope \citep[e.g.][]{10.1093/pasj/psx066,2025MNRAS.540..594K,koda,Greco2018,2023ApJ...955L..18L}, and the Dark Energy Camera Legacy Surveys (DECaLS; including Systematically Measuring Ultra Diffuse Galaxies SMUDGes: \citealt{Zaritsky2019}) as well as the Dragonfly Telephoto Array \citep[e.g.][]{2014PASP..126...55A}. In addition to camera sensitivity, the detection of these diffuse stellar structures faces another technical challenge: the preservation of the LSB signal throughout the image processing pipelines. In particular, sub-optimal local background subtraction leads to an extremely problematic loss of flux, especially in the case of extended, faint objects. The aforementioned projects have therefore had to resort to the development of dedicated pipelines optimised for the preservation of LSB signal (for instance, \texttt{Elixir-LSB}: \citealt{2012ApJS..200....4F,2015MNRAS.446..120D}), whose development remains central for future surveys \citep[e.g.][]{Borlaff-EP16,2023MNRAS.520.2484K}.

The European Space Agency's \Euclid space telescope \citep{EuclidSkyOverview} will pioneer the observation of large portions of the sky from space. During its Early Release Observations (using the ERO LSB-optimised pipeline, \citealt{EROData}) and its first quick data release (Q1, \citealt{Q1-TP001}), \Euclid proved to be an efficient LSB machine, allowing teams to detect stellar structures down to the LSB regime \citep{EROPerseusOverview}, dwarf galaxies and their nuclei \citep{EROPerseusDGs,Q1-SP001}, intracluster light (ICL; e.g. \citealt{EP-Bellhouse}) and its globular clusters in the Perseus and Fornax clusters \citep{EROPerseusICL,EROFornaxGCs}, and tidal features in the Dorado group galaxies \citep{Urbano24}. Notably, the above-mentioned studies could be extended to the Euclid Wide Survey (EWS; \citealt{Scaramella-EP1}) provided that its pipeline, different from that of the ERO, preserves the LSB signal.

This paper aims to assess the performance of the EWS in detecting Local Universe LSB objects, focusing on dwarf galaxies, including ultra-diffuse and LSB galaxies, as case studies. We use ${\rm SB}_{\rm e}$, the mean surface brightness inside the effective radius, $R_{\rm e}$, defined as
\begin{equation}
{\rm SB}_{\rm e} = \IE + 2.5\log_{10}(2\pi R_{\rm e}^2)
\label{eq:surface_brightness}
\end{equation}
in magnitudes per square arcsecond, with $\IE$ as the total apparent magnitude of the dwarf in the visible imager (VIS) band and $R_{\rm e}$ in arcsec, for a galaxy with an axis ratio of $b/a=1$, as is the case for the mock dwarfs used in this paper.

We focus on results of source injection in optical \Euclid images. The structure of the paper is as follows: Sect. \ref{sc:Data_&_Methods} presents the data utilised in this study and details our methods for injecting dwarfs and the parameters studied. The detection capabilities and limits are presented in Sect. \ref{sc:Results} and followed by a discussion in Sect. \ref{sc:Discussion}. Finally, the conclusions are summarised in Sect. \ref{sc:Conclusion}.

\section{\label{sc:Data_&_Methods}Data and methods} 

In this section we present the \Euclid data and the chosen approach for injecting dwarfs into them. In particular, we give relevant details about the EWS merge processing function (referred to as the MER pipeline), whose main outputs are the MER mosaics constructed from stacked exposures and the final MER catalogues. Given the overwhelming amount of image data to be handled, studies such as the search for dwarf galaxies across the full extent of the EWS will rely on selection criteria applied to the final MER catalogue (e.g. \citealt{Q1-SP001}). In this section we detail the method we used to assess the performance of this product.

\subsection{\Euclid images and MER catalogues}

\subsubsection{\label{dataset}Definition of the dataset}

The \Euclid space telescope operates with a visible and a near-infrared instrument. This article focuses on the VIS (\citealt{Cropper16,EuclidSkyVIS}) that uses the optical $\IE$ filter with a detector composed of $6 \times 6=36$ CCDs (i.e. $4 \times 36=144$ quadrants) covering $0.57\,\text{deg}^2$ with a pixel scale of $\ang{;;0.1}$.

In the EWS data processing, the captured raw images are processed by the \Euclid Science Ground Segment (SGS) standard pipelines. The first step is the calibration through the VIS pipeline \citep{Q1-TP002}. The output are calibrated single exposure data cubes (which include debiased and flat-fielded science images with their magnitude zeropoint, flag maps, and noise maps). There are also separate files for weights and background maps generated using the \texttt{NoiseChisel} tool from \texttt{GNU Astronomy Utilities} (\texttt{Gnuastro}, \citealt{2015ApJS..220....1A,Akhlaghi19}), intended for use in exposure co-addition. This co-addition is performed by \texttt{SWarp} \citep{SWarp} in a second pipeline, MER \citep{Q1-TP004}, which creates $0.25\,\text{deg}^2$ mosaics\footnote{In other works, such an image is often referred to as a tile. Here, however, we follow the \Euclid convention used in \cite{Q1-TP004}, where a tile denotes a unit sky region processed by the MER pipeline. Each tile is covered by a set of mosaics, one in each \Euclid band.} from stacked images, subtracts the VIS background, calculates and subtracts an additional local background, and produces the source catalogues using \texttt{SourceXtractor++} \citep{2020ASPC..527..461B,Borlaff-EP16}. Typically, each patch of the sky is covered by four long and two short VIS single calibrated exposures of a \Euclid Reference Observation Sequence (ROS). A mosaic can require more exposures to be fully covered, since depending on its location, a tile can cover areas of several observations.

Since the start of the EWS data acquisition, SGS pipeline products\footnote{For more information on the VIS and MER data models, we refer to the data product description page: \url{http://st-dm.pages.euclid-sgs.uk/data-product-doc/dmq1/}.
It should be noted that the data model and the MER pipeline are constantly evolving. In this work, we use the data model\,10 and the version 11.2 of the MER pipeline from February 2025, which is representative of the Q1 processing.} have been distributed on the fly to the Euclid Consortium through the ESA Science Archive. From those data, we extracted a set of 16 EWS calibrated single exposures in $\IE$ that suffice to produce one complete MER mosaic. The chosen MER tile (see Appendix \ref{App_tiles}) is centred on $\text{RA}=\ra{02;35;43.41}, \text{Dec}=\ang{-51;30;00.00}$ or $l=271.148\degree, b=-58.706\degree$. It corresponds to an expected background level of approximately $22\,\text{mag\,arcsec}^{-2}$ in $\IE$, which is typical of the EWS (see background estimations of \citealt{Scaramella-EP1,Borlaff-EP16}, which takes into account the zodiacal light, the Milky Way interstellar medium, and the cosmic infrared background).

\subsubsection{Running the \Euclid MER pipeline}

The numerical tools required to run the SGS pipelines are grouped under the \Euclid development environment. While EWS data are processed in dedicated data centres, for the specific needs of this study we used a \texttt{Docker} container equipped with this environment in order to execute each step of the MER pipeline locally. We categorise those steps in several groups listed below, though we refer to \cite{Q1-TP004} for a more detailed description.

\begin{itemize}
\item Stacking and tiling with VIS background subtraction.
\item Final mosaics production with MER additional local background subtraction.
\item Detection, segmentation, and de-blending of the objects.
\item Input point spread functions (PSFs) co-addition and selection of the reference PSF for each source based on its position.
\item Photometry computation and morphology characterisation for each source.
\item Sérsic fitting on the source cutouts.
\item Production of final MER catalogues combining the photometry, morphology, PSF, and Sérsic fitting information for each source.
\end{itemize}

This work includes running the MER pipeline on the set of VIS calibrated single exposures defined in Sect. \ref{dataset}. The final MER VIS mosaic delivered by the pipeline and the final MER catalogue are extracted from the outputs. The VIS \texttt{NoiseChisel} background and the MER background are subtracted from this product.

\subsection{Mock dwarf galaxies}

To assess the capabilities of the MER pipeline for detecting dwarf galaxies, we inject such objects into single exposures, run the MER pipeline and verify if the injected objects appear in the final MER catalogue with the right parameters. This requires building a mock dwarf galaxy catalogue with realistic parameters and a dwarf injection routine.

\subsubsection{Simulated dwarf galaxy parameters}

\begin{figure}[h!]
\centering
\includegraphics[width=\linewidth]{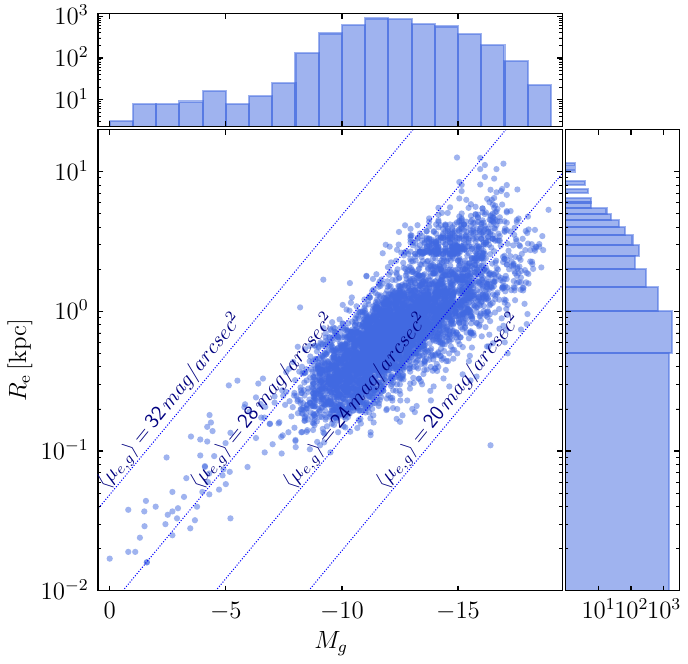}
\caption{Scaling relation between the effective radius and absolute magnitude in the $g$-band of the dwarfs galaxies in the reference sample. We indicate the average surface brightness within $R_{\rm e}$ in the $g$-band with dotted lines.}
\label{fig:Re_vs_Mg}
\end{figure}

The goal is to assemble a rather complete sample dwarf galaxies reported so far in the literature to estimate the capability of \Euclid to recover them.
We gathered catalogues of dwarf galaxies -- including ultra-faint dwarfs and ultra-diffuse galaxies -- at distance up to 120\,Mpc, ensuring comprehensive coverage across the range of simulation distance (10\,Mpc, 20\,Mpc, 70\,Mpc, and 100\,Mpc), selected according to the distances explored in the ERO. The galaxies are located in all types of environments from the field to galaxy groups and clusters. We summarise the used catalogues together with their respective environment and distance in Table \ref{tab:catalogs}. Combining all the catalogues, we obtain a reference sample of 4861 galaxies. Based on available morphological information, the reference sample contains about 83\% of dwarf ellipticals, and some irregulars show a mixed morphology with both star formation clumps and a diffuse component. Only models of galaxies corresponding to dwarf ellipticals are injected in the simulations, as this morphological type has been dominating dwarf galaxy surveys, and such galaxies can serve as proxies for the diffuse component of late-type dwarfs. In Fig.\,\ref{fig:Re_vs_Mg}, we show their size-luminosity relation.

\begin{table}
\small
\begin{center}
\caption{\label{tab:catalogs}Reference sample for simulated dwarf galaxies.}
\begin{tabular}{ccc}
\toprule
 Environment & Distance & References \\
  & [Mpc] & \\
\toprule
Local Group    & < 3 & \citet{McConnachie2012};\\
 & & \citet{Simon2019}\\
Local Volume   & < 12 & \citet{Carlsten2020}\\
Field and group & 10--120 &  \citet{Merritt2016};\\
 & & \citet{Leisman2017};\\
 & & \citet{Roman2017b};\\
 & & \citet{Greco2018};\\
 & & \citet{Forbes2020};\\
 & & \citet{2021MNRAS.506.5494P};\\
 & & \citet{Marleau2021}\\
Virgo cluster & 16.5 & \citet{Ferrarese2020};\\
 & & \citet{Lim2020}\\
Fornax cluster & 20 & \citet{Eigenthaler2018};\\
 & & \citet{Venhola2017};\\
 & & \citet{venhola2022}\\
Hydra I cluster & 51 & \citet{Iodice2020}\\
Perseus cluster & 72 & \citet{EROPerseusDGs} \\
Coma cluster & 100 & \citet{Zaritsky2019}\\
\bottomrule
\end{tabular}
\end{center}
\end{table}

The dwarf galaxies injected into the VIS images were modelled using a Sérsic profile \citep{1963BAAA....6...41S}. For each galaxy, we set the Sérsic index to 0.8, i.e. the median value of the reference sample, as our study is focusing on the effect on the magnitude and effective radius ($R_{\rm e}$) of the galaxies only\footnote{The impact of varying the Sérsic index between 0.7 and 1 was reported to have a negligible effect on the modelling of UDGs in \citet{Leisman2017}, which gives us confidence in our results for Sérsic indices reasonably close to 0.8. Nevertheless, it is worth noting that this study does not cover more extreme (and thus rarer) values of the Sérsic index, for which the results may differ.}.
We randomly select the absolute magnitude and $R_{\rm e}$ in kpc by picking a galaxy from the reference sample. The drawn $R_{\rm e}$ and absolute magnitude $M_g$ are then converted to arcsecond and apparent magnitude according to the simulation distance ($D$). The dwarfs are set to be round. At each $D$, we draw 100 galaxies considering only dwarfs at a similar distance or smaller to also test the detection of close-by faint ones, such as ultra-faint dwarfs in the Local Group.

\begin{figure}[h!]
\centering
\includegraphics[width=\linewidth]{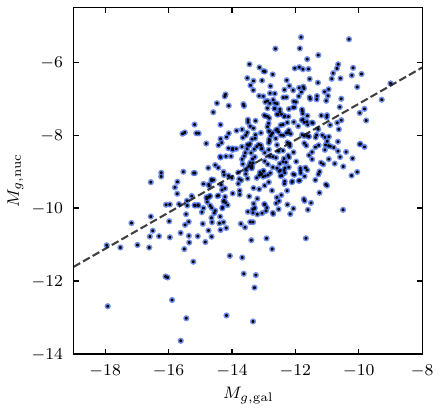}
\caption{Relation between the absolute magnitude in the $g$-band of the nucleus and dwarf galaxy host for nucleated dwarf galaxies in the Virgo and Fornax clusters as well as in the field and galaxy groups. The linear fit is shown with a dashed line.}
\label{fig:Mnuc_vs_Mgal}
\end{figure}

To investigate the effect of the presence of a nucleus on the detection of the galaxy, we clone each simulated dwarf and model a nucleus at the centre by using a King profile \citep{1966AJ.....71...64K}, leading to nearly 200 galaxies per simulation (i.e. 800 galaxies in total). Following the simulations of bright globular clusters from \citet{EP-Voggel}, we modelled a nucleus such that FWHM\,=\,4\,pc. The magnitude of the nucleus is determined from the linear relation $M_{g\rm ,nuc} = 0.5M_{g\rm ,gal}-2.17$ between the galaxy and nucleus absolute magnitude shown in Fig.\,\ref{fig:Mnuc_vs_Mgal}, fitted on the properties of nucleated dwarfs from \cite{Sanchez-Janssen2019} in the Virgo galaxy cluster, \cite{Eigenthaler2018} and \cite{Ordenes2018} in the Fornax cluster, and \cite{2021MNRAS.506.5494P} in field and galaxy groups environment.

\begin{figure*}
\centering
\includegraphics[trim={5cm 0.5cm 5cm 0cm},clip,width=\linewidth]{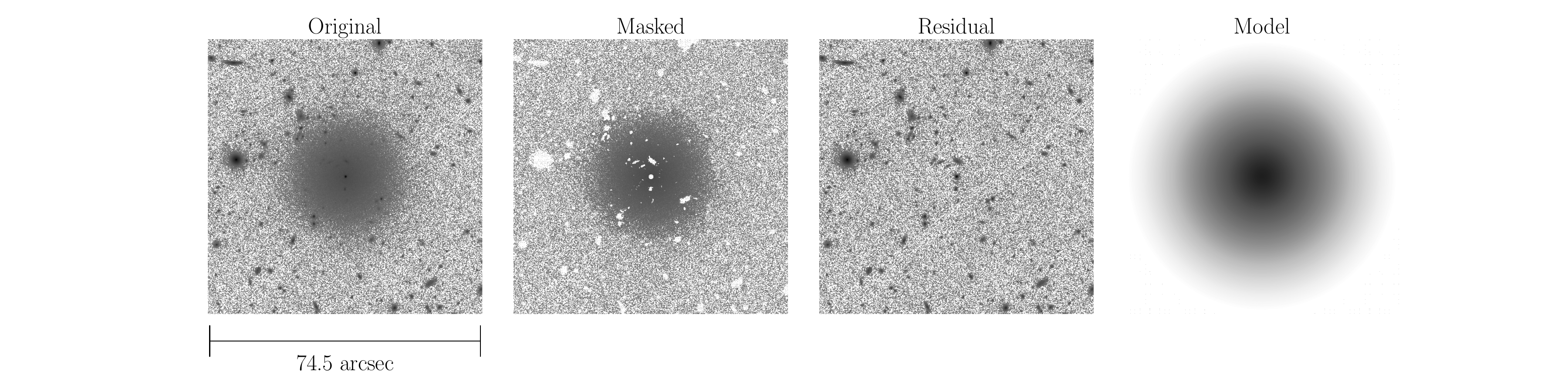}
\caption{Summary of the \texttt{Galfit} fitting strategy, including (from left to right) the original image (i.e. the cutout of an injected dwarf galaxy at 10\,Mpc), the masked image obtained with a combination of \texttt{MTObjects} and \texttt{Source Extractor} segmentation maps, the residual image obtained by subtraction of the original image with the model, and the model image.}
\label{fig:galfit_strategy}
\end{figure*}

The position of each dwarf on the exposures is defined in order to simulate about 200 objects at each distance. This number is chosen to obtain reliable statistics without overcrowding the exposures. To avoid galaxy overlap, we define a square area of side from $9\,R_{\rm e}$ to $42\,R_{\rm e}$ around each dwarf where no other dwarf is injected. The smallest area ensures having enough sky to model the galaxy (as explained in Sect. \ref{sec:visual_inspection}), whereas areas larger than $9\,R_{\rm e}$ are defined as a function of $D$ so that $\sim$200 objects are distributed all over the exposure.

\subsubsection{Injecting mock dwarf galaxies in \Euclid calibrated single exposures}

We develop the \texttt{Python} code \texttt{LSBSim} to inject dwarfs and their nuclei in real EWS images. For a given \Euclid VIS calibrated single exposure, \texttt{LSBSim} receives a list of galactic parameters (right ascension, declination, total magnitude, position angle, effective radius, ellipticity, and Sérsic index). It has also as input a list of nuclear star cluster parameters (right ascension, declination, total magnitude, ellipticity, core radius, and tidal radius). Then, \texttt{LSBSim} scans the 144 quadrants to inject these nearby objects. The program uses stamps where galaxies are generated by the \texttt{GalSim} Python package \citep{2014ascl.soft02009R} and then injected at the specified world coordinate system location on the different quadrants. We convolve the injected objects with the \Euclid VIS PSF (calculated from ERO data in \citealt{Urbano24,Saifollahi25}, see also \citealt{EROData}) and add Poisson noise to these stamps before the injection. The native \texttt{GalSim} package contains a Sérsic function used for generating dwarf galaxies. For the nuclei, we developed a function to allow \texttt{GalSim} to handle the King model.

This procedure allows us to update the single exposure data cubes with the injected objects, more specifically the science image and the Poisson-noise root mean square map, quadrant by quadrant, with the injected Local Universe objects. We needed to update the background files corresponding to those single exposures as well. To do this, we ran \texttt{NoiseChisel} on each quadrant containing the injected Local Universe objects using exactly the same configuration parameters as in the VIS pipeline. This resulted in a complete set of MER pipeline input for each exposure. 

After the MER pipeline run, we extracted the MER mosaics before and after background subtraction and the final MER catalogues. Those products are analysed in the following sections.

\subsection{\label{detect_fit}Sérsic model fitting with \texttt{Galfit}}

In the MER catalogue, the Sérsic parameters of galaxies were estimated by fitting each source cutout with \texttt{SourceXtractor++} runs included in the pipeline. We compare the obtained parameters with the injected ones and with those we measured using the method detailed in the next paragraphs.

We produced a cutout in the final MER mosaic for each of the injected dwarfs with a side length of nine effective radii, as suggested in \cite{2021MNRAS.506.5494P} to encompass enough background around the dwarf galaxy. We used the \texttt{Galfit} \citep{2002AJ....124..266P} software to fit the dwarfs on each of our cutouts. In particular, we used initial parameters close to the input ones, and we also fitted a tilted plane background with zero as the initial level entered in the software.
This approach requires masking all sources except for the object to be fitted. However, the number of injected dwarf galaxies does not allow for manual masking of each of the associated cutouts. As in \cite{EROPerseusDGs}, we used a combination of segmentation maps produced by \texttt{MTObjects} \citep{teeninga2015improved} and \texttt{Source\,Extractor} \citep{1996A&AS..117..393B}, from which we removed the central detection to prevent the galaxy of interest from being masked. We masked a small circular region at the centre of each galaxy, taking into account the presence of a nucleus and its size. This combination brings out the best in both software packages, respectively providing generous mask sizes for extended objects and sensitivity to compact sources. This strategy is exemplified in Fig.\,\ref{fig:galfit_strategy}.

\subsection{\label{sec:visual_inspection}Visual inspection}

Visual inspection is used to assess the detectability of the injected mock dwarf galaxies in the images. One expert inspects each cutout, classifying the dwarfs into two categories. Dwarfs were classified as `detected by eye' if the object was visually identified as a dwarf and as `undetected' if the object's nature was uncertain or there was no visual detection. The human eye still outperforms segmentation algorithms in detecting and identifying dwarf galaxies, even with a small number of classifiers and when these algorithms are optimised for LSB (see, for instance, a performance comparison between visual inspection and \texttt{MTObjects} in \citealt{2020A&A...644A..91M}).

\begin{figure*}
\centering
\includegraphics[width=0.495\linewidth]{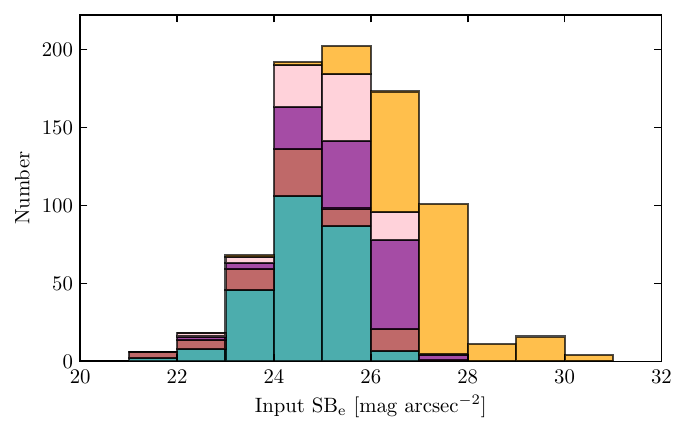}
\includegraphics[width=0.495\linewidth]{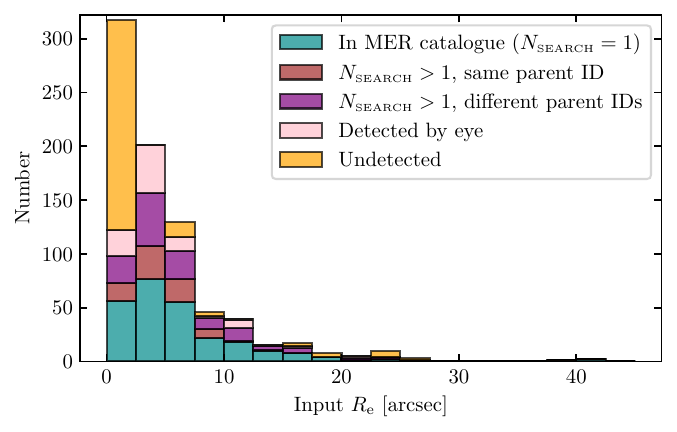}
\vskip -0.3cm
\caption{Histograms of the input ${\rm SB}_{\rm e}$ (left panel) and the input $R_{\rm e}$ (right panel), colour-coded according to their detection by eye and in the MER catalogue. Here, $N_\sfont{SEARCH}$ is the number of MER sources found by using the search radius $R_{\sfont{SEARCH}}$ for the cross-match. It is worth noting that the dwarfs in MER catalogues and those with $N_\sfont{SEARCH}>1$ are also detected by eye. In this plot, all the dwarfs (nucleated or not) are included. In complement of this plot, Appendix \ref{AppDetectionRate} provides a ${\rm SB}_{\rm e}$-$R_{\rm e}$ map of detection rate. Finally, in Appendix \ref{AppA}, we also provide the histogram of $\langle \mu_\sfont{I} \rangle$ as defined in \cite{Q1-SP001}.}
\label{fig:detect_catalog_histo}
\end{figure*}

\begin{table*}[h!]
\caption{Comparison between the detection statistics across different ${\rm SB}_{\rm e}$ bins.}
\centering
\renewcommand{\arraystretch}{1.2}
\setlength{\extrarowheight}{2pt}
\begin{tabular}{ccccccc}
\hline \hline
\noalign{\vskip 4pt}
${\rm SB}_{\rm e}$ $[\text{mag\,arcsec}^{-2}]$ & Input dwarfs & In MER catalogue & Single parent ID & Single or multiple parent ID(s) & Detected by eye \\
(1) & (2) & (3) & (4) & (5) & (6) \\
\noalign{\vskip 4pt}
\hline
\noalign{\vskip 4pt}
21–22 & $6\,(100\,\%)$ & $2\,(33\,\%)$ & $6\,(100\,\%)$ & $6\,(100\,\%)$ & $6\,(100\,\%)$ \\
22–23 & $18\,(100\,\%)$ & $8\,(44\,\%)$ & $14\,(78\,\%)$ & $16\,(89\,\%)$ & $18\,(100\,\%)$ \\
23–24 & $68\,(100\,\%)$ & $46\,(68\,\%)$ & $59\,(87\,\%)$ & $63\,(93\,\%)$ & $67\,(99\,\%)$ \\
24–25 & $193\,(100\,\%)$ & $106\,(55\,\%)$ & $136\,(70\,\%)$ & $163\,(84\,\%)$ & $190\,(98\,\%)$ \\
25–26 & $202\,(100\%)$ & $87\,(43\%)$ & $98\,(49\,\%)$ & $141\,(70\,\%)$ & $184\,(91\,\%)$ \\
26–27 & $173\,(100\,\%)$ & $7\,(4\%)$ & $21\,(12\,\%)$ & $78\,(45\,\%)$ & $96\,(55\,\%)$ \\
27–28 & $101\,(100\,\%)$ & $0\,(0\%)$ & $1\,(1\,\%)$ & $4\,(4\,\%)$ & $5\,(5\,\%)$ \\
28–29 & $11\,(100\,\%)$ & $0\,(0\,\%)$ & $0\,(0\,\%)$ & $0\,(0\,\%)$ & $0\,(0\,\%)$ \\
29–30 & $16\,(100\,\%)$ & $0\,(0\,\%)$ & $0\,(0\,\%)$ & $0\,(0\,\%)$ & $0\,(0\,\%)$ \\
30–31 & $4\,(100\,\%)$ & $0\,(0\,\%)$ & $0\,(0\,\%)$ & $0\,(0\,\%)$ & $0\,(0\,\%)$ \\
\noalign{\vskip 4pt}
\hline \hline
\end{tabular}
\tablefoot{Column (1) gives the ${\rm SB}_{\rm e}$ bin. Columns (2) indicates the number of injected dwarfs. The remaining columns are the number of dwarfs that are (3) present in the final MER catalogue as one single object, (4) present in the final MER catalogue as one or several objects sharing the same parent ID, (5) present in the final MER catalogue as one or several objects sharing or not the same parent ID(s), and (6) visually detected, whether or not they are present in the final MER catalogue. As a result, columns (3) to (5) are cumulative. Columns (2) to (6) are given in number and in percent of the injected dwarfs in the corresponding surface brightness bin. In this table, all the dwarfs (nucleated or not) are included.}
\label{tab:sb_bins}
\end{table*}

\begin{table*}[h!]
\caption{Similar to Table \ref{tab:sb_bins}, but now binning the data in $R_{\rm e}$ instead of in ${\rm SB}_{\rm e}$.}
\centering
\renewcommand{\arraystretch}{1.2}
\setlength{\extrarowheight}{2pt}
\begin{tabular}{ccccccc}
\hline \hline
\noalign{\vskip 4pt}
$R_{\rm e}$ [arcsec] & Input dwarfs & In MER catalogue & Single parent ID & Single or multiple parent ID(s) & Detected by eye \\
(1) & (2) & (3) & (4) & (5) & (6) \\
\noalign{\vskip 4pt}
\hline
\noalign{\vskip 4pt}
0–2.5 & $317\,(100\,\%)$ & $56\,(18\,\%)$ & $73\,(23\,\%)$ & $98\,(31\,\%)$ & $122\,(38\,\%)$ \\
2.5–5 & $201\,(100\,\%)$ & $77\,(38\,\%)$ & $107\,(53\,\%)$ & $157\,(78\,\%)$ & $201\,(100\,\%)$ \\
5–7.5 & $130\,(100\,\%)$ & $55\,(42\,\%)$ & $77\,(59\,\%)$ & $103\,(79\,\%)$ & $116\,(89\,\%)$ \\
7.5–10 & $46\,(100\,\%)$ & $22\,(48\,\%)$ & $30\,(65\,\%)$ & $40\,(87\,\%)$ & $42\,(91\,\%)$ \\
10–12.5 & $39\,(100\%)$ & $18\,(46\%)$ & $19\,(49\,\%)$ & $31\,(79\,\%)$ & $39\,(100\,\%)$ \\
12.5–15 & $15\,(100\,\%)$ & $10\,(67\%)$ & $11\,(74\,\%)$ & $14\,(93\,\%)$ & $15\,(100\,\%)$ \\
15–17.5 & $17\,(100\,\%)$ & $8\,(47\%)$ & $8\,(47\,\%)$ & $13\,(76\,\%)$ & $14\,(82\,\%)$ \\
17.5–20 & $8\,(100\,\%)$ & $4\,(50\,\%)$ & $4\,(50\,\%)$ & $4\,(50\,\%)$ & $4\,(50\,\%)$ \\
20–22.5 & $5\,(100\,\%)$ & $1\,(20\,\%)$ & $1\,(20\,\%)$ & $3\,(60\,\%)$ & $5\,(100\,\%)$ \\
22.5–25 & $10\,(100\,\%)$ & $2\,(20\,\%)$ & $2\,(20\,\%)$ & $4\,(40\,\%)$ & $4\,(40\,\%)$ \\
25–27.5 & $3\,(100\,\%)$ & $0\,(0\,\%)$ & $0\,(0\,\%)$ & $1\,(33\,\%)$ & $1\,(33\,\%)$ \\
37.5–40 & $1\,(100\,\%)$ & $1\,(100\,\%)$ & $1\,(100\,\%)$ & $1\,(100\,\%)$ & $1\,(100\,\%)$ \\
40–42.5 & $2\,(100\,\%)$ & $2\,(100\,\%)$ & $2\,(100\,\%)$ & $2\,(100\,\%)$ & $2\,(100\,\%)$ \\
\noalign{\vskip 4pt}
\hline \hline
\end{tabular}
\label{tab:re_bins}
\end{table*}

\section{\label{sc:Results}Results}

\subsection{\label{detection_mer}Dwarf galaxy detection}

In this subsection we assess the capabilities and limitations of MER for detecting Local Universe dwarf galaxies. We cross-matched the injected dwarf coordinates with the final MER catalogues (using a search radius $R_{\sfont{SEARCH}}$, which is typically $R_{\rm e}$ and is further discussed in Appendix \ref{App_rsearch}). We find, in particular, a completeness of 91\,\% for ${\rm SB}_{\rm e}\in[21,24]$, and 54\,\% for ${\rm SB}_{\rm e}\in[24,28]$ in mag\,arcsec$^{-2}$ (see Appendix \ref{App_non_corrected_det} for full results). These values represent upper limits and do not take into account any contaminant correction, including confusion with background galaxies.

The final corrected results are shown in Fig.\,\ref{fig:detect_catalog_histo} and detailed below. These histograms illustrate the detected fraction of dwarf galaxies, depending on ${\rm SB}_{\rm e}$ and $R_{\rm e}$. These fractions in each bin are reported in Tables \ref{tab:sb_bins} and \ref{tab:re_bins}, respectively. We complete the analysis by separating the dwarfs based on their distance, and reproducing the same plots in Appendix \ref{AppB}. We distinguish four possible cases regarding the recovery of injected dwarf galaxies in the MER catalogue.

\begin{figure}[h!]
\centering
\includegraphics[width=\linewidth]{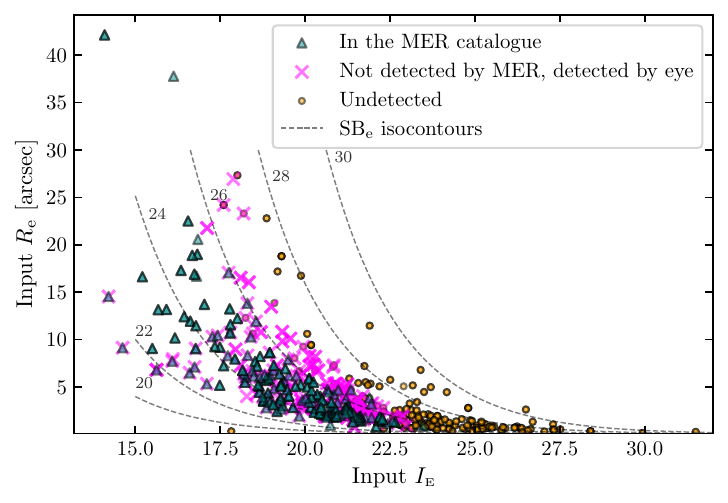}
\vskip -0.3cm
\caption{Input $R_{\rm e}$ as a function of the total magnitude in $\IE$ used to inject the dwarfs. They are colour-coded according to their detection (in the final MER catalogue, only by eye or not detected).}
\label{fig:detect_catalog_params}
\end{figure}

In a first case, the injected dwarf has one match in the final MER catalogue within $R_{\sfont{SEARCH}}$. To visualise the characteristics of such objects, we plot the parameter space (i.e. the input $R_{\rm e}$ as a function of the input total magnitude $\IE$) in Fig.\,\ref{fig:detect_catalog_params}. We observe that, for a given magnitude, the final MER catalogue misses the most extended galaxies. The region of the parameter space beyond 22.5 as total input magnitude in $\IE$ is dominated by objects that are not detected. Combining all ${\rm SB}_{\rm e}$ bins up to 28\,mag\,arcsec$^{-2}$, $34\,\%$ of the injected dwarf galaxies are recovered as single sources in the final MER catalogue. For ${\rm SB}_{\rm e}<24\,\text{mag\,arcsec}^{-2}$ (regular dwarf regime), $61\,\%$ of the injected dwarfs are recovered in the final MER catalogue. For ${\rm SB}_{\rm e}>24\,\text{mag\,arcsec}^{-2}$ (LSB regime), $30\,\%$ of the injected dwarfs are recovered in the final MER catalogue.

\begin{figure}[h!]
\centering
\includegraphics[width=\linewidth]{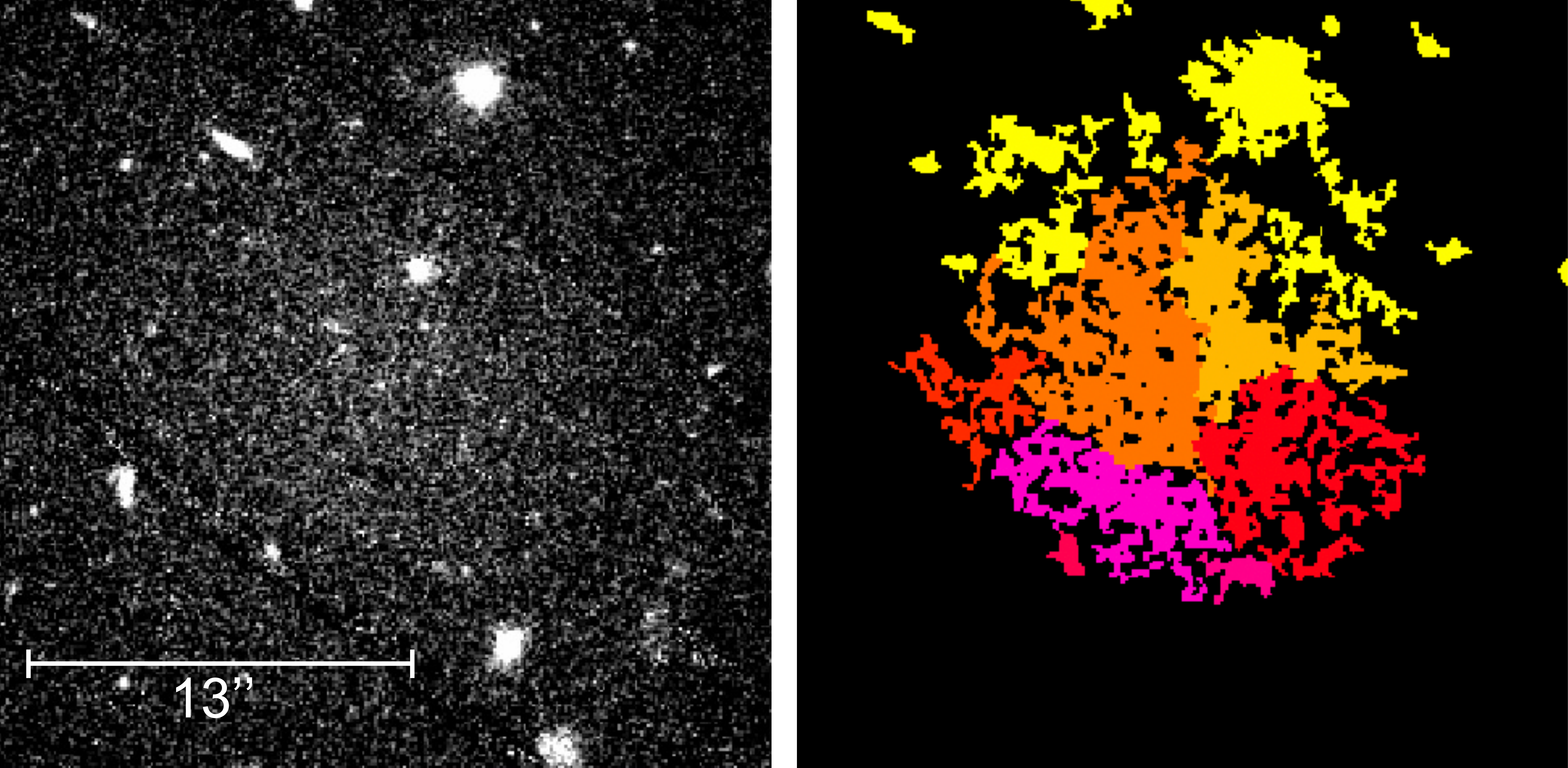}
\caption{Example of segmentation and de-blending for an injected dwarf galaxy (left panel). The associated mask (right panel) is divided into multiple sources, differentiated by different colours.}
\label{fig:segmentation}
\end{figure}

In a second case, the injected dwarf is associated with multiple MER sources with different IDs within $R_{\sfont{SEARCH}}$. In such cases, the dwarf appears fragmented into several areas in the segmentation map generated during the MER pipeline run (see an example in Fig.\,\ref{fig:segmentation}). This effect arises from its de-blending step. The MER pipeline assigns the same value in the \texttt{PARENT\_ID} column of the final MER catalogue (hereafter referred to as the `parent ID') to all sources that were initially segmented together but later separated during de-blending. We can then distinguish between two subtypes of detection:

\begin{itemize}
\item Single parent ID detection: The dwarf galaxy was correctly segmented but split into multiple sources during de-blending. In this case, all regions making up the dwarf galaxy share the same parent ID, and it is possible to regroup them afterwards to measure the properties of the object. Combining all ${\rm SB}_{\rm e}$ bins up to 28\,mag\,arcsec$^{-2}$, $44\,\%$ of the injected dwarf galaxies are recovered as single sources or multiple sources with the same parent ID in the final MER catalogue. For ${\rm SB}_{\rm e}<24\,\text{mag\,arcsec}^{-2}$ (regular dwarf regime), $86\,\%$ of the injected dwarfs are recovered in the final MER catalogue. For ${\rm SB}_{\rm e}>24\,\text{mag\,arcsec}^{-2}$ (LSB regime), $38\,\%$ of the injected dwarfs are recovered in the final MER catalogue.
\item Multiple parent IDs: the faint dwarf is segmented into multiple regions assigned to different parent IDs even before de-blending. Such objects are difficult to recover afterwards.
\end{itemize}

In a third case, the dwarf has MER detections within $R_{\sfont{SEARCH}}$, but these correspond to background sources and\,/\,or the dwarf cannot be identified as such. Visual inspection, along with a cross-match between the input dwarf catalogues and the MER mosaic generated from a pipeline run without any injected dwarfs, ensures that such cases are not counted among the detected dwarfs.

\begin{figure}[h!]
\centering
\includegraphics[width=\linewidth]{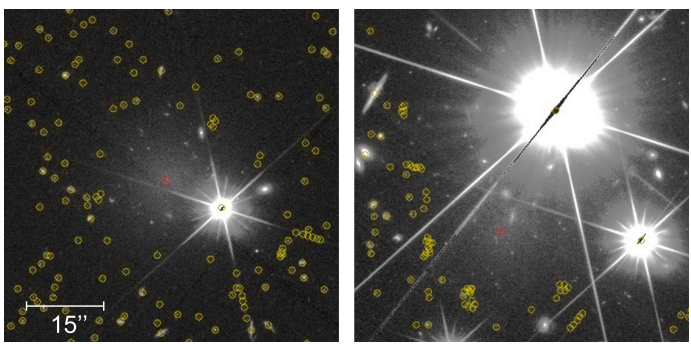}
\caption{Examples of non-detection of a dwarf galaxy by the MER pipeline in the vicinity of a Milky Way star. The MER pipeline detections are labelled with blue circles, the injected dwarfs are labelled with red squares.}
\label{fig:stars}
\end{figure}

The fourth case is the absence of detection in the MER catalogue. This occurs either when the dwarf is too faint to be detected and there are no background sources within $R_{\sfont{SEARCH}}$ or when its mask is mistakenly merged with that of a nearby bright extended object (including, but not limited to, Milky Way stars, as exemplified in Fig.\,\ref{fig:stars}).

The detection limit of the catalogue (cyan + brown bars in Fig.\,\ref{fig:detect_catalog_histo}), as well as that of the visual detection (purple + rose bars in Fig.\,\ref{fig:detect_catalog_histo}) is between ${\rm SB}_{\rm e}=27\,\text{mag\,arcsec}^{-2}$ and ${\rm SB}_{\rm e}=28\,\text{mag\,arcsec}^{-2}$ (corresponding to a signal-to-noise ratio between 2 and 3). Together with Fig. \ref{fig:detect_catalog_params}, we can also highlight $\IE=22.5$ as the threshold distinguishing a regime dominated by detected dwarf galaxies from another regime dominated by dwarf galaxies which are not. Below $R_{\rm e}\approx3\arcsec$, very few detections are made.
Note that most of these small and LSB dwarfs can be referred to as ultra-faint dwarfs, which we do not expect to detect beyond 20\,Mpc.

We explored whether the initial VIS background subtraction using \texttt{NoiseChisel} or the subsequent MER local background subtraction have an impact on the detection of dwarf galaxies. To do so, we compared the final MER mosaic, the mosaic prior to the MER background subtraction but after the VIS background subtraction (`VIS BGSUB mosaic' hereafter), and the mosaic before both background subtractions as a reference point (`NOBG mosaic' hereafter). We extracted the VIS BGSUB mosaic as an intermediate product produced during the MER pipeline run. The NOBG mosaic is not a standard product delivered during the MER pipeline run. However, it can be easily generated with a second run of the pipeline using constant background files instead of the VIS background files and by extracting the mosaic before the second (local MER) background subtraction. The visual inspection for the final mosaic was repeated for the `VIS BGSUB mosaic' and the `NOBG mosaic' (see Appendix \ref{Detectability_nuclei}). This analysis showed that the background has no significant impact on the dwarf detection.

\begin{figure}[h!]
\centering
\includegraphics[trim={1.2cm 0cm 0cm 0cm},clip,width=\linewidth]{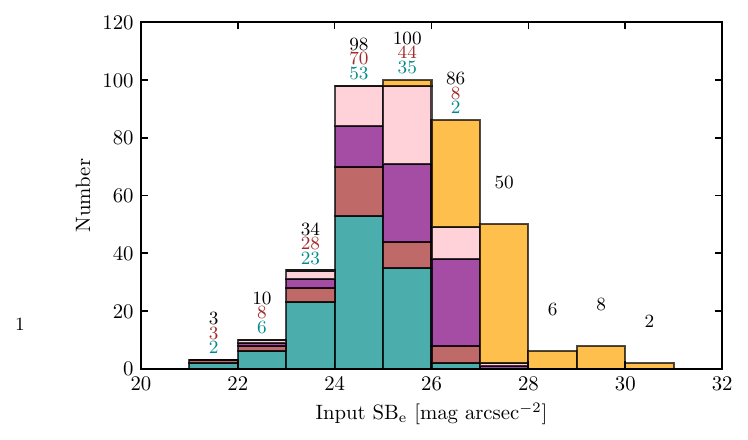}
\vskip -0.3cm
\caption{Identical to the left panel of Fig. \ref{fig:detect_catalog_histo}, but only for the non-nucleated dwarfs. Above each bin, the number of injected dwarfs is shown in black, the number of dwarfs recovered as either a single MER source or as multiple fragments with the same parent ID is shown in brown, and the number recovered as a single MER source is shown in cyan.}
\label{fig:no_nuclei_detection_histo}
\end{figure}

\begin{figure}[h!]
\centering
\includegraphics[trim={1.2cm 0cm 0cm 0cm},clip,width=\linewidth]{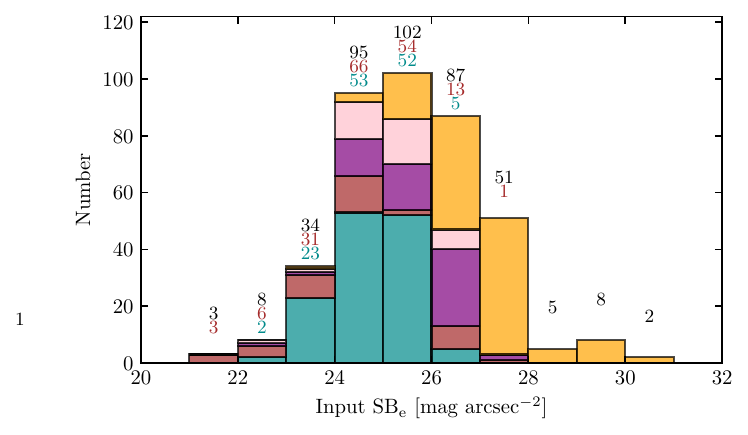}
\vskip -0.3cm
\caption{Identical to Fig. \ref{fig:no_nuclei_detection_histo}, but for nucleated dwarfs.}
\label{fig:nuclei_detection_histo}
\end{figure}

Finally, we also investigated the impact of the presence or absence of a nucleus on detection, by reproducing in Fig.\,\ref{fig:no_nuclei_detection_histo} and Fig.\,\ref{fig:nuclei_detection_histo} the histogram of the input ${\rm SB}_{\rm e}$ from Fig.\,\ref{fig:detect_catalog_histo}, this time separating nucleated dwarfs from non-nucleated ones. Detection slightly favours nucleated dwarfs. When summing all surface brightness bins up to 28\,mag\,arcsec$^{-2}$, 42\,\% of the injected non-nucleated dwarfs are recovered, compared to 46\,\% for nucleated dwarfs. Specifically, in the non-LSB regime, 83\,\% of non-nucleated dwarfs are detected versus 89\,\% of nucleated ones. In the LSB regime, 37\,\% of non-nucleated dwarfs are detected compared to 40\,\% for nucleated dwarfs. The largest difference is found in the 25--26\,mag\,arcsec$^{-2}$ bin, amounting to 9\,\%.

\subsection{\label{dwarf_parameters}Dwarf galaxy parameter measurements}

The MER pipeline was developed with a strong focus on high-quality photometry for compact sources, such as distant galaxies, in order to meet the requirements of cosmology, the core science of the \Euclid mission. In this subsection we examine the parameters returned by the final MER catalogue for diffuse sources in the Local Universe and assess the ability of the pipeline to support science for which it was not originally optimised. Then, we explore the impact of the different background subtractions.

\begin{figure}[h!]
\centering
\includegraphics[width=\linewidth]{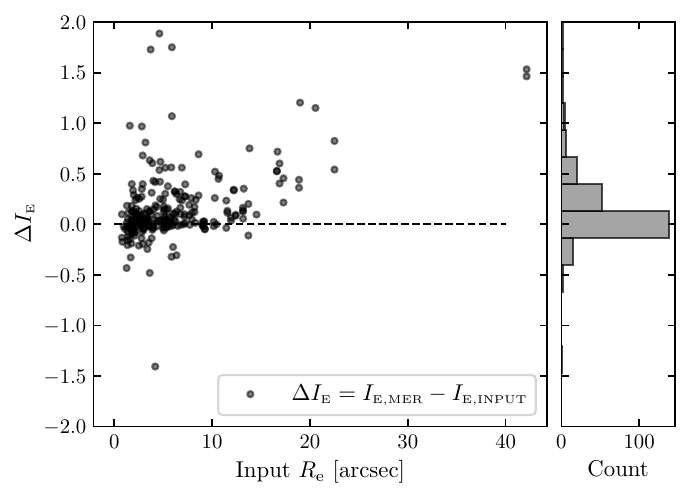}
\vskip -0.3cm
\caption{Difference between the \texttt{SourceXtractor++} measured magnitude in the final MER catalogue and the input magnitude of injected dwarfs as a function of $R_{\rm e}$ for all dwarfs detected in the final MER catalogue.}
\label{fig:detect_catalog_2}
\end{figure}

\subsubsection{Parameters in output of the MER pipeline}

Based on the cross-match between the input catalogue of injected galaxies and the list of dwarfs detected as a single source in the final MER catalogue, we can compute the difference between the measured magnitude (derived from the MER catalogue column \texttt{FLUX\_VIS\_SERSIC}) and the input magnitude as a function of $R_{\rm e}$ (Fig.\,\ref{fig:detect_catalog_2}). We observe two main effects: a scatter and a flux loss which increases with the radius.

Regarding the scatter in the recovered magnitude, we interpret cases where the flux is overestimated as instances where bright objects (including, but not exclusively, overlapping stars) remain within the detected galaxy. To explain the more common cases of flux underestimation, we note that the MER segmentation masks may not be extended enough to encompass most of the flux of the dwarf galaxy. This can lead to the production of cutouts that are too small for a proper fitting by \texttt{SourceXtractor++}.

Regarding the flux loss increasing with the radius, this happens beyond an effective radius of $10\,\arcsecond$ (it reaches $\Delta\IE\approx1.5$ at $R_{\rm e}\approx40\arcsec$). Such an effect may indicate a local background oversubtraction. This becomes problematic when the object for which we aim to measure photometry exceeds the size of a cell used to estimate the background. The impact of the different background subtractions applied during the MER pipeline run is explored in the following subsections. We repeat the model fitting with \texttt{Galfit} described in Sect. \ref{detect_fit} for the final MER mosaic is repeated for the `VIS BGSUB mosaic' and the `NOBG mosaic'.

\begin{figure*}[h!]
    \centering
    \begin{subfigure}{0.4905\textwidth}
        \centering
        \includegraphics[trim={0cm 0cm 0cm 0cm},clip,width=\linewidth]{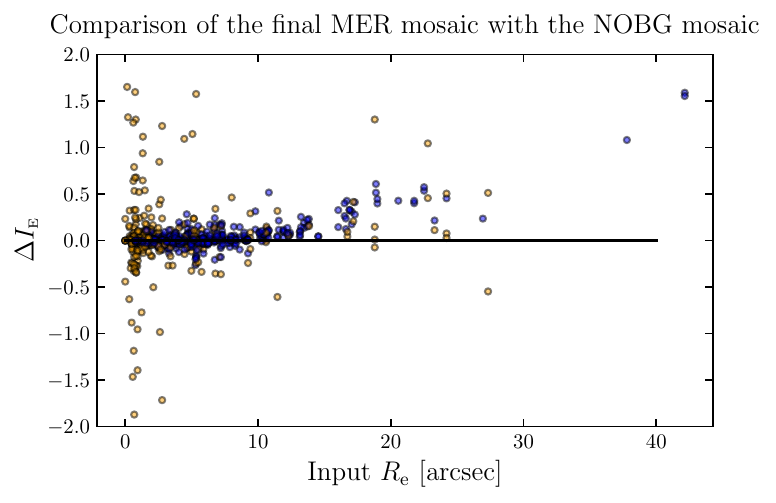}  
    \end{subfigure}
    \begin{subfigure}{0.498\textwidth}
        \centering
        \includegraphics[width=\linewidth]{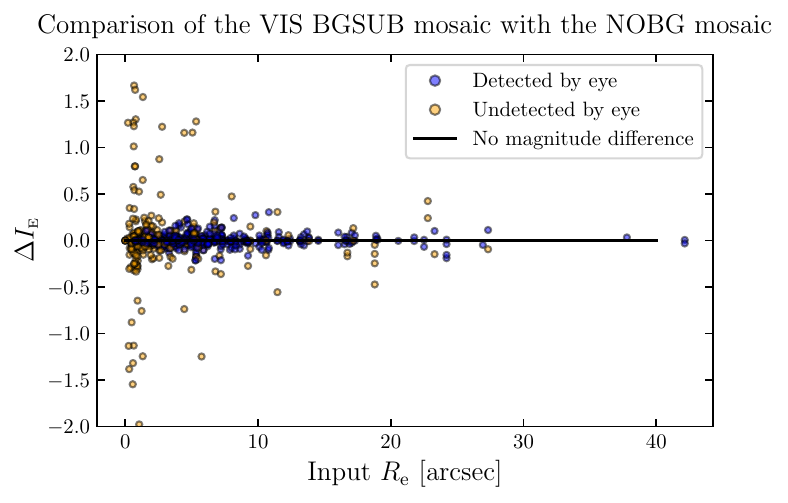}  
    \end{subfigure}
    \caption{Magnitude difference $\Delta\IE=I_{\sfont{E,BGSUB}}-I_{\sfont{E,NOBG}}$ as a function of the input effective radius for the final MER mosaic and the VIS BGSUB mosaic cases. The total magnitudes were obtained using \texttt{Galfit}.}
    \label{fig:diff}
\end{figure*}

\begin{figure}[h!]
\centering
\includegraphics[width=\linewidth]{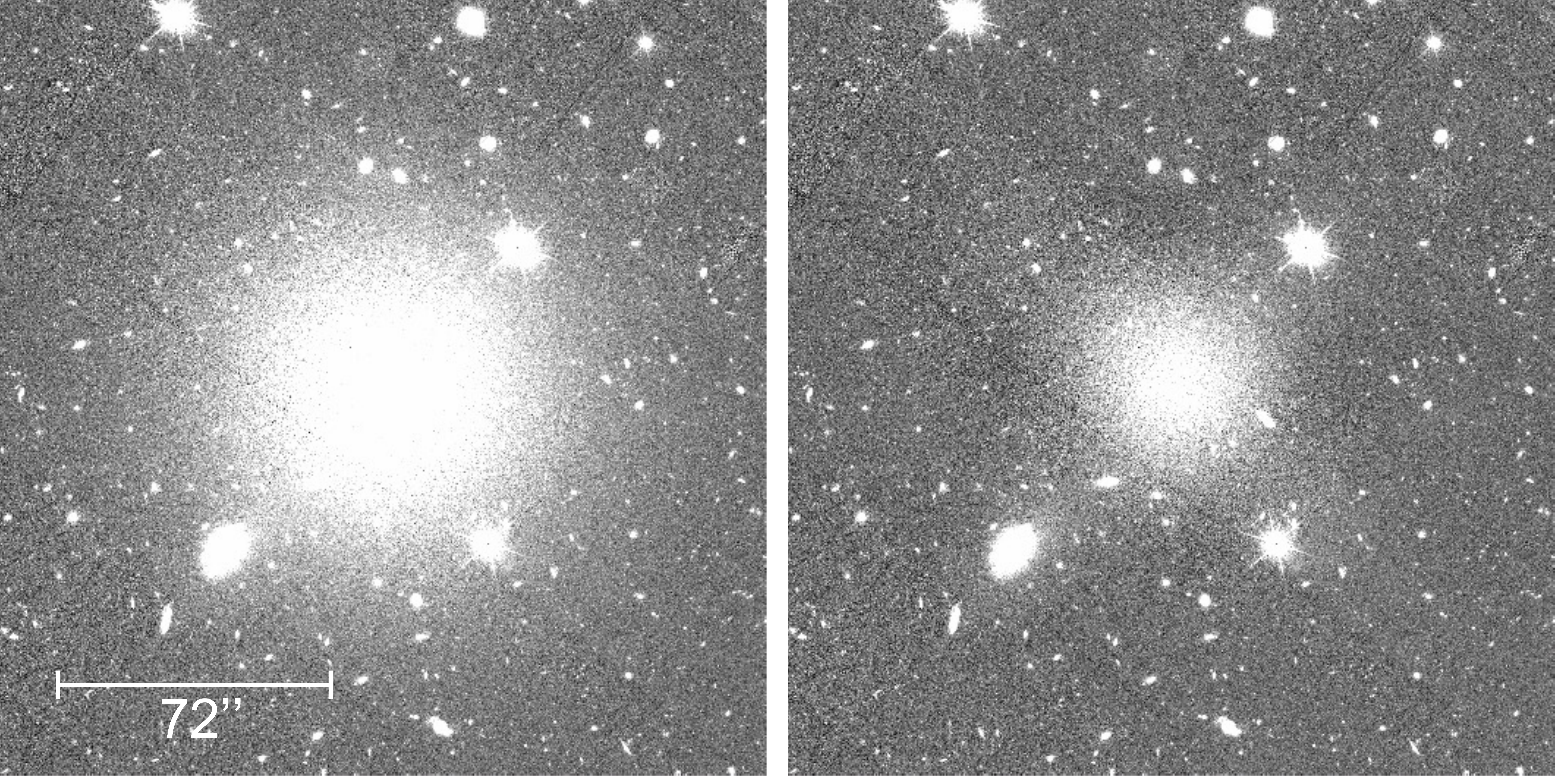}
\caption{Example of a dwarf galaxy successfully detected in the VIS BGSUB mosaic (left panel). The second background subtraction affects its appearance (along with the square-shaped background mesh used in the process) and consequently its structural parameters in the final MER mosaic (right panel). Both images use the same scale. It is worth noting that this is the brightest ($\IE=18$) and most extended galaxy in our sample at 10\,Mpc and that such objects are statistically rare in the Local Universe.}
\label{fig:params}
\end{figure}

\subsubsection{\label{MER_bkg}Impact of the MER background}

To further test the impact of the MER background subtraction, we compared the final fully background-subtracted mosaic with the intermediate product where only the \texttt{NoiseChisel} VIS background has been applied (no local MER background subtraction). Figure \ref{fig:diff} displays the impact of each background subtraction on the \texttt{Galfit} total $\IE$ magnitude ($\Delta\IE=I_{\sfont{E}\text{,BGSUB}}-I_{\sfont{E}\text{,NOBG}}$) as a function of the input $R_{\rm e}$. The large scatter at very small $R_{\rm e}$ corresponds to the regime in which dwarf galaxy identification and model fitting become challenging, and where the PSF may significantly affect the derived parameters. The right panel shows the impact of the MER local background subtraction. As also observed in Fig.\,\ref{fig:detect_catalog_2}, a difference in the magnitudes is visible beyond $R_{\rm e}=10\arcsecond$, with the deviation increasing until reaching 1.5 magnitude at $R_{\rm e}\approx40\arcsecond$. Thus, after the MER local background subtraction, the structural parameters of the dwarfs with $R_{\rm e}\geq10\arcsecond$ are modified. An example of a dwarf galaxy whose structural parameters are significantly affected after the second background subtraction is presented in Fig.\,\ref{fig:params}.

\subsubsection{Impact of the VIS background}

We tested the effect of the initial \texttt{NoiseChisel} VIS background subtracted mosaic by comparing it to the custom mosaic where no background subtraction is applied. To check whether the initial \texttt{NoiseChisel} background subtraction impacts the measured parameters of the dwarf galaxies, we perform the same test as before by comparing the magnitudes measured on the images without any background subtraction and after this first background subtraction (respectively the `NOBG mosaic' and the `VIS BGSUB mosaic'). This is what is shown on the left panel of Fig.\,\ref{fig:diff}. We observe that the trend observed for galaxies detected by eye corresponds to a null magnitude difference. Thus, this first background subtraction does not impact the structural parameters of the detected dwarfs.

\section{\label{sc:Discussion}Discussion}

\subsection{Making the most of MER pipeline products}

Provided we are able to merge the dwarf galaxy fragments that share the same parent ID (a step that must be carried out after the MER pipeline run, since the pipeline itself neither performs this operation nor outputs the parameters of the re-merged sources), the final MER catalogue is nearly 90\,\% complete down to a surface brightness of $24\,\text{mag\,arcsec}^{-2}$. The final MER catalogue parameters remain reliable for galaxies with a size up to $R_{\rm e} = 10\arcsec$, beyond which a flux loss is observed. The analysis and comparison of the final MER mosaic, the VIS BGSUB mosaic, and the NOBG mosaic allowed us to observe that this flux loss emerges with the subtraction of the second, MER local background. Indeed, we do not observe any changes in the dwarf galaxy parameters when subtracting only the VIS background, but we observe the flux loss when subtracting the MER background from the VIS background subtracted image.

The final MER catalogue is derived from the final MER mosaic which is MER background subtracted (i.e. the parameters of the dwarf galaxies before the local background subtraction are not measured and thus are not available in the MER catalogues). The safest way to recover the parameters of Local Universe dwarf galaxies (particularly beyond $R_{\rm e} = 10\arcsec$) is then to run outside of the MER pipeline a model fitting program on the final mosaic or its cutouts, after having first re-added the second MER local background (thereby reverting to the VIS BGSUB mosaic, subtracted only from the VIS background). This can be easily achieved using the background maps (`BGMOD') provided in the ESA Science Archive.

It is important to highlight that the VIS background is estimated at the quadrant scale (approximately $3\arcmin\times3\arcmin$), which could lead to flux loss for objects approaching or exceeding this size. However, most dwarf galaxies in the Local Universe do not reach this size. This suggests that the VIS background subtracted products from the \Euclid SGS pipelines are fully compliant with the science of most dwarfs. This result cannot be extended to studies concerning the extended haloes of giant galaxies, including their tidal features, and intracluster light. Indeed, they are typically more extended and thus are more likely to be affected by this quadrant scale limitation.

A detection approach relying solely on the MER catalogues would miss the most extended and faint dwarf galaxies (notably more than 50\,\% of those between 24 and 28\,mag\,arcsec$^{-2}$ in effective surface brightness, beyond which dwarfs are no longer identifiable in \Euclid images) as well as those located near bright and\,/\,or extended objects, which tend to be jointly segmented.

To overcome some of the limitations mentioned above, one may propose possible modifications to the SGS pipelines. One possible solution to make the SGS pipelines LSB-compliant at the scale of a full \Euclid field of view (FoV) would be to replace the current quadrant-by-quadrant VIS background estimation with a FoV-scale approach (that is, first aligning the background levels of each quadrant, then applying \texttt{NoiseChisel} to a full VIS exposure composed of all quadrants, rather than processing each quadrant independently). To avoid flux loss in the output MER catalogues, an additional object parameter measurement step would need to be introduced in the MER pipeline before the local background subtraction (preferably using an LSB-optimised segmentation and de-blending tools such as \texttt{MTObjects}). In the course of this catalogue production procedure, it would be desirable to reduce the area around bright stars where sources cannot be detected. This can be achieved using the same LSB-optimised segmentation and de-blending algorithms mentioned earlier.

\subsection{Limitations of this study}

We expect that several studies will build upon this work. Two promising directions for future work are the inclusion of colour information and the exploration of a wider range of injected dwarf galaxy morphologies. This subsection outlines these two avenues.

For this paper, we have considered only the $\IE$ band, as its greater depth and higher resolution compared to the NIR bands make it the optimal detection band in \Euclid. This is especially true in the Local Universe, where dwarf galaxies that are undetected in the $\IE$ band are not expected to be detectable in any of the NIR bands. Nevertheless, a new study will be required to test the impact of background subtraction on the colours of dwarf galaxies. Indeed, not only can the second background subtraction performed by MER alter the measured colour of Local Universe objects, but a first source of error may arise from the initial background subtraction, which is calculated differently in the VIS and NIR pipelines (especially, in the NIR pipeline, the background is computed at the scale of each of its 16 single $10\,\arcmin\times10\,\arcmin$ detectors; see also \citealt{Q1-TP003}). Including the study of the NIR bands would also allow for assessment of how colour selection can improve the detection and identification of dwarf galaxies (in particular distant ones, which cannot be distinguished from background sources using the $\IE$ image alone).

Also, in this paper, we have only injected elliptical galaxies, as they represent the most common type among the real dwarf catalogues used to find the mock dwarf parameters. We also limited ourselves to selecting galaxies already identified as dwarfs in those catalogues, which, by definition, led us to omit dwarfs with rarer morphologies that may not have been classified as such. Indeed, our goal here was to cover a realistic parameter space for dwarf galaxies. However, to diversify the nature of the injected dwarfs, it will then be necessary to consider also more elongated dwarfs (which present additional detection challenges, \citealt{2023ApJ...955L..18L}) and use more realistic dwarf galaxy models (including for dwarf irregulars), likely derived from simulations (e.g. mock images extracted from \texttt{IllustrisTNG}, presented in \citealt{Nelson2018TheIS}). This will allow for better determination of the multi-band detection capabilities and limits for dwarfs with different morphological types and star-formation histories. The inclusion of dwarf galaxies with star-formation clumps will also help assess the fragmentation of such objects into multiple sources (linked or not by the same parent ID) by MER.

Finally, it is worth noting that the results presented in this paper are more applicable to dwarf galaxies in the field, in the sense that each mock galaxy is sufficiently isolated from its neighbours so that its light does not affect their detection or parameter fitting (as is often the case in the field). The effect of dwarf galaxy clustering will be addressed in a future paper focusing on the injection of mock dwarf galaxies in different environments (including galaxy clusters).

\subsection{Comparison to other \Euclid works}

Several other \Euclid papers also discuss the detection of dwarf galaxies, such as \cite{Q1-SP001} and \cite{EROPerseusOverview}, respectively in the context of Q1 and ERO Perseus data. One of the detection limits highlighted in these two studies, as well as in ours, is the total magnitude detection threshold.

In our paper, we have highlighted $\IE=22.5$ as the threshold distinguishing a regime dominated by detected dwarf galaxies from another regime dominated by dwarf galaxies which are not. \cite{Q1-SP001} report detections up to about 1 magnitude fainter, based on measurements from the MER catalogue. This can be explained, on the one hand, by the use of colour images from \Euclid, which may facilitate the identification of small and\,/\,or faint dwarfs and their distinction from background sources. On the other hand, it may be due to less reliable photometry in the MER catalogue for these faint objects (as seen in Fig. \ref{fig:detect_catalog_2}, a 1 magnitude difference between the real and measured total magnitude is possible).

Finally, the objective of our paper is to assess the capability of \Euclid standard pipeline products to probe Local Universe dwarf galaxies. A different study, which is beyond the scope of this paper, will consist in the injection of a large number of dwarfs at various distances. It would be necessary to establish predictions on the detection of the faint end of the dwarf galaxy luminosity function as a function of redshift and environment. Such a work has already been initiated for \Euclid images in the specific case of estimating the galaxy luminosity function in the ERO Perseus data. This study, described in Appendix B of \cite{EROPerseusOverview}, reveals as well a break and a drop in completeness at $\IE=22.5$ (see Fig. B.6 of this paper).

\subsection{Prospects and generalisation to other wide surveys}

Using the MER catalogues could be an efficient way to build a training set for machine learning and deep learning algorithms, in order to automatically identify numerous dwarf galaxies in \Euclid images. However, it should be noted that in a training sample solely based on MER catalogue detections, LSB dwarfs will be underrepresented.

Investigating the effects of background subtraction from the \Euclid pipeline serves as a valuable test-bed for other wide surveys, as it employs two widely used subtraction methods: one designed to be LSB-compliant (\texttt{NoiseChisel}) and another optimised for compact source analysis (local background subtraction, here applied through \texttt{SourceXtractor++}). Several current and future wide surveys use a local background subtraction, such as the DESI Legacy imaging surveys (e.g. the Dark Energy Camera Legacy Survey DECaLS, \citealt{2019AJ....157..168D}), while other are still developing this step of their image processing pipeline, as is the case for the Vera Rubin Observatory's Legacy Survey of Space and Time \citep{2019ApJ...873..111I,2020arXiv200111067B,2024MNRAS.528.4289W}. Such consideration of the treatment of the background is also relevant for missions on longer timescales (e.g. \textit{Nancy Roman Grace}: \citealt{2023AAS...24210102K} and ARRAKIHS: \citealt{2024eas..conf.1990G}). Our study demonstrates that only the first method reliably preserves Local Universe dwarf parameters.

\section{\label{sc:Conclusion}Conclusions}

Achieving high completeness in the detection of dwarf galaxies down to the LSB regime is crucial for studies of galaxy evolution and near-field cosmology. The wide survey of the \Euclid Space Telescope will unveil the LSB Universe with high resolution and deep VIS and NIR imaging over a large portion of the extragalactic sky, as demonstrated by early works on its first images. In this study, we have measured the dwarf detection ability of \Euclid by the injection of mock dwarfs and nuclei into individual exposures. We ran the EWS MER pipeline and analysed its products in the VIS band. The MER pipeline includes background subtractions at different scales: one at the scale of a VIS quadrant and another at a local scale. We investigated the impact of these subtractions on the detection and parameter measurement of dwarf galaxies. We conclude regarding the MER products in VIS, whose images and catalogues are usually used for dwarf galaxy detection:

\begin{itemize}
\item Although the MER pipeline is fine-tuned for cosmology rather than LSB science, its final catalogues exhibit a high level of completeness for dwarf galaxies with ${\rm SB}_{\rm e}\leq24\,\text{mag\,arcsec}^{-2}$ ($86\,\%$). Beyond this surface brightness, the MER catalogues still successfully recover a fraction of the injected dwarf galaxies ($38\,\%$).

\item The background subtractions performed by the MER pipeline do not affect the detections.

\item The local background subtraction causes a flux loss in dwarf galaxies larger than $R_{\rm e}=10\arcsecond$. Re-adding this background (which is easily achievable using the final mosaics and the background maps provided by MER) and then performing parameter measurements on these mosaics is sufficient to correct this issue. The extracted dwarf parameters remain accurate until the galaxy reaches arcminute scales, with the upper limit corresponding to that of a VIS quadrant (i.e. $3\arcmin$). Typically, Local Universe dwarf galaxies do not reach this size.
\end{itemize}

We note that including the NIR bands will make it possible to assess the impact of colour on the identification of dwarf galaxies. A future study will include this analysis as well as the injection of more complex and realistic dwarf galaxies.

Identifying all LSB dwarfs that are intrinsically detectable by \Euclid as well as studying Local Universe objects more extended than $3\arcmin$ (typically giant galaxies, their tidal features, and ICL) require replacing the background subtractions applied in the MER pipeline with an alternative post-processing of the calibrated VIS single exposures, which are fully LSB compliant. The results described in this article help advance understanding of background subtraction effects of both LSB-compliant methods and local background subtraction methods, thus providing useful insight for current and upcoming deep surveys.

\begin{acknowledgements}
The authors thank the anonymous referee for their constructive report. The authors thank Thomas Oliveira, Nicolas Mai and Samuel Rusterucci for insightful  discussions. Junais is funded by the European Union (MSCA EDUCADO, GA 101119830 and WIDERA ExGal-Twin, GA 101158446). JHK acknowledges grant PID2022-136505NB-I00 funded by
MCIN/AEI/10.13039/501100011033 and EU, ERDF. The Euclid Consortium acknowledges the European Space Agency and a number of agencies and institutes that have supported the development of Euclid, in particular the Agenzia Spaziale Italiana, the Austrian Forschungsförderungsgesellschaft funded through BMK, the Belgian Science Policy, the Canadian Euclid Consortium, the Deutsches Zentrum für Luft- und Raumfahrt, the DTU Space and the Niels Bohr Institute in Denmark, the French Centre National d’Etudes Spatiales, the Fundação para a Ciência e a Tecnologia, the Hungarian Academy of Sciences, the Ministerio de Ciencia, Innovación y Universidades, the National Aeronautics and Space Administration, the National Astronomical Observatory of Japan, the Netherlandse Onderzoekschool Voor Astronomie, the Norwegian Space Agency, the Research Council of Finland, the Romanian Space Agency, the State Secretariat for Education, Research, and Innovation (SERI) at the Swiss Space Office (SSO), and the United Kingdom Space Agency. A complete and detailed list is available on the Euclid web site (http://www.euclid-ec.org).
M.P. and A.V. are supported by the Academy of Finland grant No. 347089.
\end{acknowledgements}

\bibliography{biblio}

\appendix

\section{\label{App_tiles}MER mosaic}

\begin{figure}[h!]
\centering
\includegraphics[width=\linewidth]{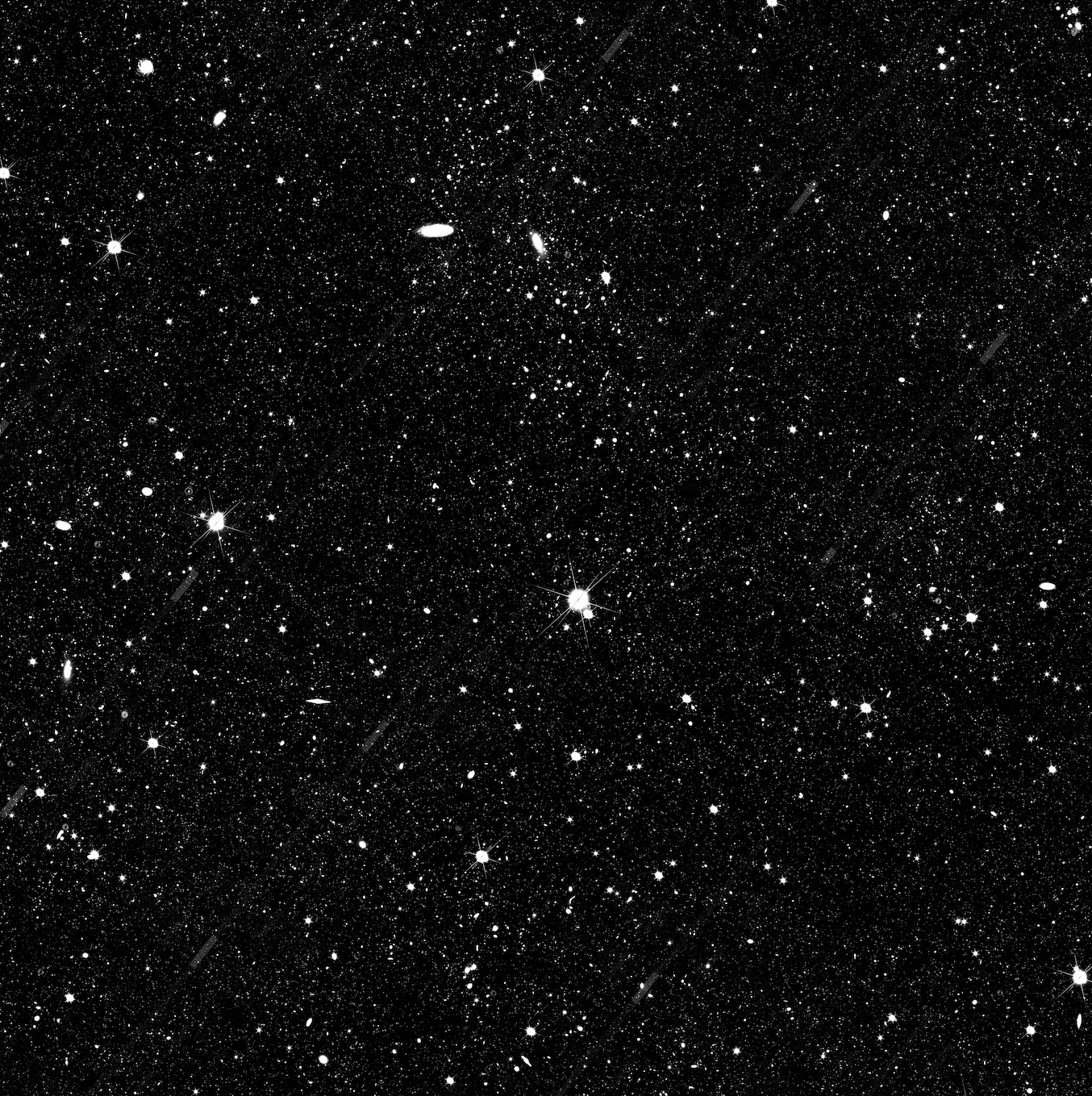}
\caption{Chosen MER VIS mosaic. The tile with ID $102018187$ is centred on ${\rm RA} = \ra{02;35;43.41}$, ${\rm Dec} = \ang{-51;30;00.00}$. Its $x$- and $y$-axes correspond to RA and Dec, respectively. Its side length is $32\arcmin$ \citep[for a description of the tiling scheme see][]{Q1-TP004}. Data from this tile will be publicly available as part of the \Euclid first data release (DR1).}
\label{fig:MER_tile}
\end{figure}

The MER VIS mosaic used for this work is  shown in Fig. \ref{fig:MER_tile}.

\section{\label{App_rsearch}Cross-match search radius}

In this appendix we discuss briefly the definition of our search radius $R_{\sfont{SEARCH}}$, used for the cross-match between the input parameter catalogue of injected mock dwarfs, and the final MER catalogue. While standard cross-match procedures use a fixed radius, we opted instead for a radius that varies depending on the galaxy being searched for. Specifically, for each dwarf in the input catalogue, we perform a search in the final MER catalogue within a $R_{\sfont{SEARCH}}$ centred on its input position, $R_{\sfont{SEARCH}}$ depending on the input size of the galaxy. We use the full input $R_{\rm{e}}$ when it is unaffected by the MER background subtraction (i.e. for $R_{\sfont{SEARCH}}<10\arcsecond$), and $R_{\rm{e}}/2$ in all other cases.

Our tests also included other $R_{\sfont{SEARCH}}$ definitions, such as the theoretical 25\,mag\,arcsec$^{-2}$ isophotal radius, which on average matched the size of the MER segmentation masks better than the effective radius, but was not defined for galaxies fainter than this surface brightness. Except for those faint galaxies, the results did not vary with the final definition adopted for $R_{\sfont{SEARCH}}$.

\newpage

\section{\label{AppDetectionRate}Detection rate map}

\begin{figure}[h!]
\centering
\includegraphics[width=\linewidth]{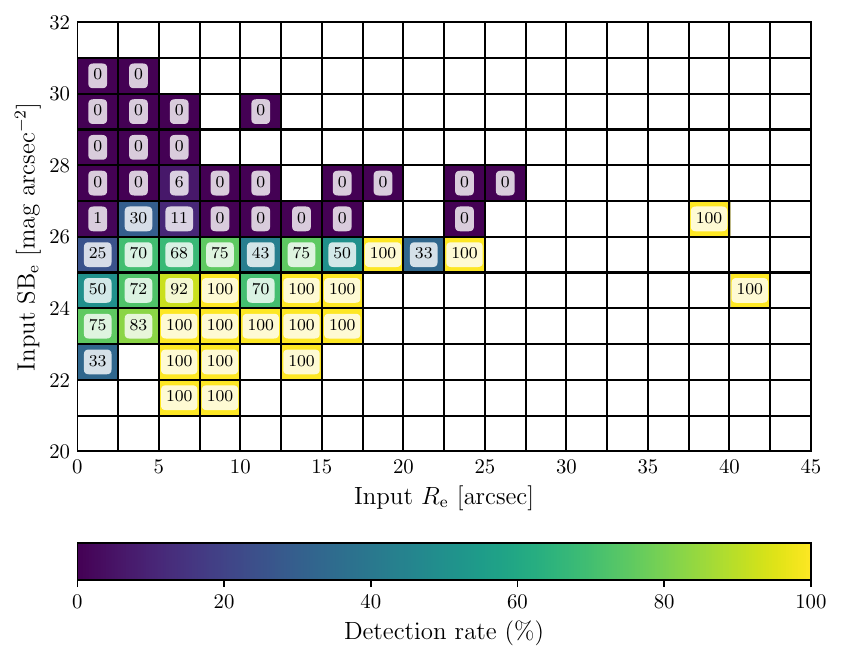}
\vskip -0.3cm
\caption{Two-dimensional histogram of the detection rate (in \%) per bin of ${\rm SB}_{\rm e}$ and $R_{\rm{e}}$. In this plot, a dwarf is defined as detected if it is a single source in the MER catalogue or if the corresponding MER catalogue sources are linked by the same parent ID.}
\label{fig:detection_rate}
\end{figure}

As a complement of the Fig. \ref{fig:detect_catalog_histo}, the detection rate of injected mock dwarf galaxies in the ${\rm SB}_{\rm e}$-$R_{\rm{e}}$ map are provided in Fig. \ref{fig:detection_rate}. It is worth noting that the comparison with completeness maps from other papers (e.g. \citealt{carlsten2022,2023ApJ...955L..18L}) is not direct. Indeed, completeness studies typically involve the injection of a very large number of sources across a wide range of sizes and luminosities (independently of the realism of such source) in order to uniformly populate the full parameter space under study. The approach chosen in this paper was, on the contrary, to restrict the analysis to already-detected dwarf galaxies, in order to assess the performance and limitations of the \Euclid pipeline specifically on such sources. As a result, the parameter space is not uniformly filled (i.e. the number of dwarfs per bin varies, since they were randomly selected from their distribution in the observations).

\FloatBarrier

\section{\label{AppA}Detectability as a function of $\langle \mu_\sfont{I} \rangle$}

\begin{figure}[h!]
\centering
\includegraphics[width=\linewidth]{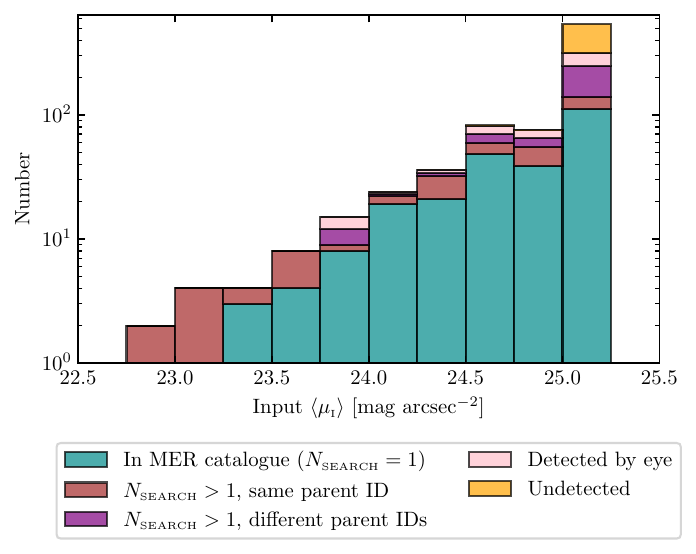}
\vskip -0.3cm
\caption{Histograms of the input $\langle \mu_\sfont{I} \rangle$, colour-coded according to
their detection by eye and in the MER catalogue, with $N_\sfont{SEARCH}$ the number of MER sources found in the search radius $R_{\sfont{SEARCH}}$ used for the cross-match.}
\label{fig:detect_catalog_histo_mu}
\end{figure}

\cite{Q1-SP001} uses $\langle \mu_\sfont{I} \rangle$, defined as the average surface brightness computed with the columns \texttt{FLUX\_SEGMENTATION} and \texttt{SEGMENTATION\_AREA} of the final MER catalogues. We thus present Fig.\,\ref{fig:detect_catalog_histo_mu}, which is a version of the Fig.\,\ref{fig:detect_catalog_histo} using these metrics. Please note that the relation between the input ${\rm SB}_{\rm e}$ of injected galaxies and the MER output $\langle \mu_\sfont{I} \rangle$ is extrapolated for the dwarfs not detected by MER.

\FloatBarrier

\section{\label{App_non_corrected_det}Non-corrected detectability}

\begin{figure}[h!]
\centering
\includegraphics[width=\linewidth]{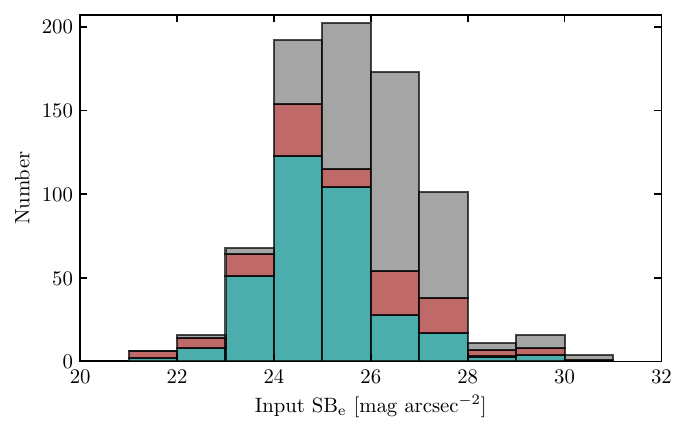}
\includegraphics[width=\linewidth]{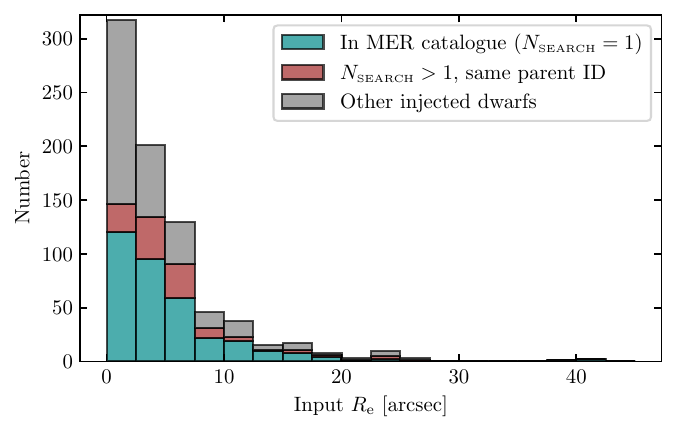}
\vskip -0.3cm
\caption{Histograms of the input ${\rm SB}_{\rm e}$ (top panel) and the input $R_{\rm e}$ (bottom panel), colour-coded according to their detection by eye and in the MER catalogue, $N_\sfont{SEARCH}$ being the number of MER sources found by using the search radius $R_{\sfont{SEARCH}}$ for the cross-match.}
\label{fig:detect_catalog_histo_no_visual}
\end{figure}

The complete results for the dwarf detectability before the contaminant correction are given in in Fig.\,\ref{fig:detect_catalog_histo_no_visual} and in Tables\,\ref{tab:sb_bins_no_visual} and \ref{tab:re_bins_no_visual}. Those values represent upper limits. It is worth noting that in the main body of the paper, the values given take into account the single parent ID detections.

\setlength{\tabcolsep}{3pt}
\begin{table}[h!]
\caption{Comparison between the detection statistics across different ${\rm SB}_{\rm e}$ bins.}
\centering
\renewcommand{\arraystretch}{1.2}
\setlength{\extrarowheight}{2pt}
\begin{tabular}{ccccccc}
\hline \hline
\noalign{\vskip 4pt}
${\rm SB}_{\rm e}$ & Input dwarfs & In MER catalogue & Single parent ID \\
(1) & (2) & (3) & (4) \\
\noalign{\vskip 4pt}
\hline
\noalign{\vskip 4pt}
21–22 & $6\,(100\,\%)$ & $2\,(33\,\%)$ & $6\,(100\,\%)$ \\
22–23 & $18\,(100\,\%)$ & $8\,(44\,\%)$ & $14\,(78\,\%)$ \\
23–24 & $68\,(100\,\%)$ & $51\,(75\,\%)$ & $64\,(94\,\%)$ \\
24–25 & $192\,(100\,\%)$ & $123\,(64\,\%)$ & $154\,(80\,\%)$ \\
25–26 & $202\,(100\%)$ & $104\,(51\%)$ & $114\,(56\,\%)$ \\
26–27 & $173\,(100\,\%)$ & $28\,(16\%)$ & $54\,(31\,\%)$ \\
27–28 & $101\,(100\,\%)$ & $17\,(17\%)$ & $38\,(38\,\%)$ \\
28–29 & $11\,(100\,\%)$ & $3\,(27\,\%)$ & $7\,(63\,\%)$ \\
29–30 & $16\,(100\,\%)$ & $4\,(25\,\%)$ & $8\,(50\,\%)$ \\
30–31 & $4\,(100\,\%)$ & $1\,(25\,\%)$ & $1\,(25\,\%)$ \\
\noalign{\vskip 4pt}
\hline \hline
\end{tabular}
\tablefoot{Column (1) gives the ${\rm SB}_{\rm e}$ bin in \text{mag\,arcsec}$^{-2}$. Columns (2) indicates the number of injected dwarfs. The remaining columns are the number of dwarfs that are (3) present in the final MER catalogue as one single object, (4) present in the final MER catalogue as one or several objects sharing the same parent ID. Column (4) includes the detections from Column (3). They are given in number and in percent of the injected dwarfs in the corresponding surface brightness bin.}
\label{tab:sb_bins_no_visual}
\end{table}

\begin{table}[h!]
\caption{Similar to Table \ref{tab:sb_bins_no_visual}, but now binning the data in $R_{\rm e}$ instead of in ${\rm SB}_{\rm e}$.}
\centering
\renewcommand{\arraystretch}{1.2}
\setlength{\extrarowheight}{2pt}
\begin{tabular}{ccccccc}
\hline \hline
\noalign{\vskip 4pt}
$R_{\rm e}$ [arcsec] & Input dwarfs & In MER catalogue & Single parent ID \\
(1) & (2) & (3) & (4) \\
\noalign{\vskip 4pt}
\hline
\noalign{\vskip 4pt}
0–2.5 & $317\,(100\,\%)$ & $120\,(38\,\%)$ & $146\,(46\,\%)$ \\
2.5–5 & $201\,(100\,\%)$ & $95\,(47\,\%)$ & $134\,(67\,\%)$ \\
5–7.5 & $130\,(100\,\%)$ & $59\,(45\,\%)$ & $91\,(70\,\%)$ \\
7.5–10 & $46\,(100\,\%)$ & $22\,(48\,\%)$ & $31\,(67\,\%)$ \\
10–12.5 & $39\,(100\%)$ & $19\,(49\%)$ & $23\,(59\,\%)$ \\
12.5–15 & $15\,(100\,\%)$ & $10\,(67\%)$ & $11\,(73\,\%)$ \\
15–17.5 & $17\,(100\,\%)$ & $8\,(47\%)$ & $11\,(65\,\%)$ \\
17.5–20 & $8\,(100\,\%)$ & $4\,(50\,\%)$ & $6\,(75\,\%)$ \\
20–22.5 & $5\,(100\,\%)$ & $1\,(20\,\%)$ & $1\,(20\,\%)$ \\
22.5–25 & $10\,(100\,\%)$ & $2\,(20\,\%)$ & $5\,(50\,\%)$ \\
25–27.5 & $3\,(100\,\%)$ & $0\,(0\,\%)$ & $1\,(33\,\%)$ \\
37.5–40 & $1\,(100\,\%)$ & $1\,(100\,\%)$ & $1\,(100\,\%)$ \\
40–42.5 & $2\,(100\,\%)$ & $2\,(100\,\%)$ & $2\,(100\,\%)$ \\
\noalign{\vskip 4pt}
\hline \hline
\end{tabular}
\label{tab:re_bins_no_visual}
\end{table}

\section{\label{AppB}Distance and dwarf detectability}

\begin{figure*}[h!]
\centering
\includegraphics[width=\linewidth]{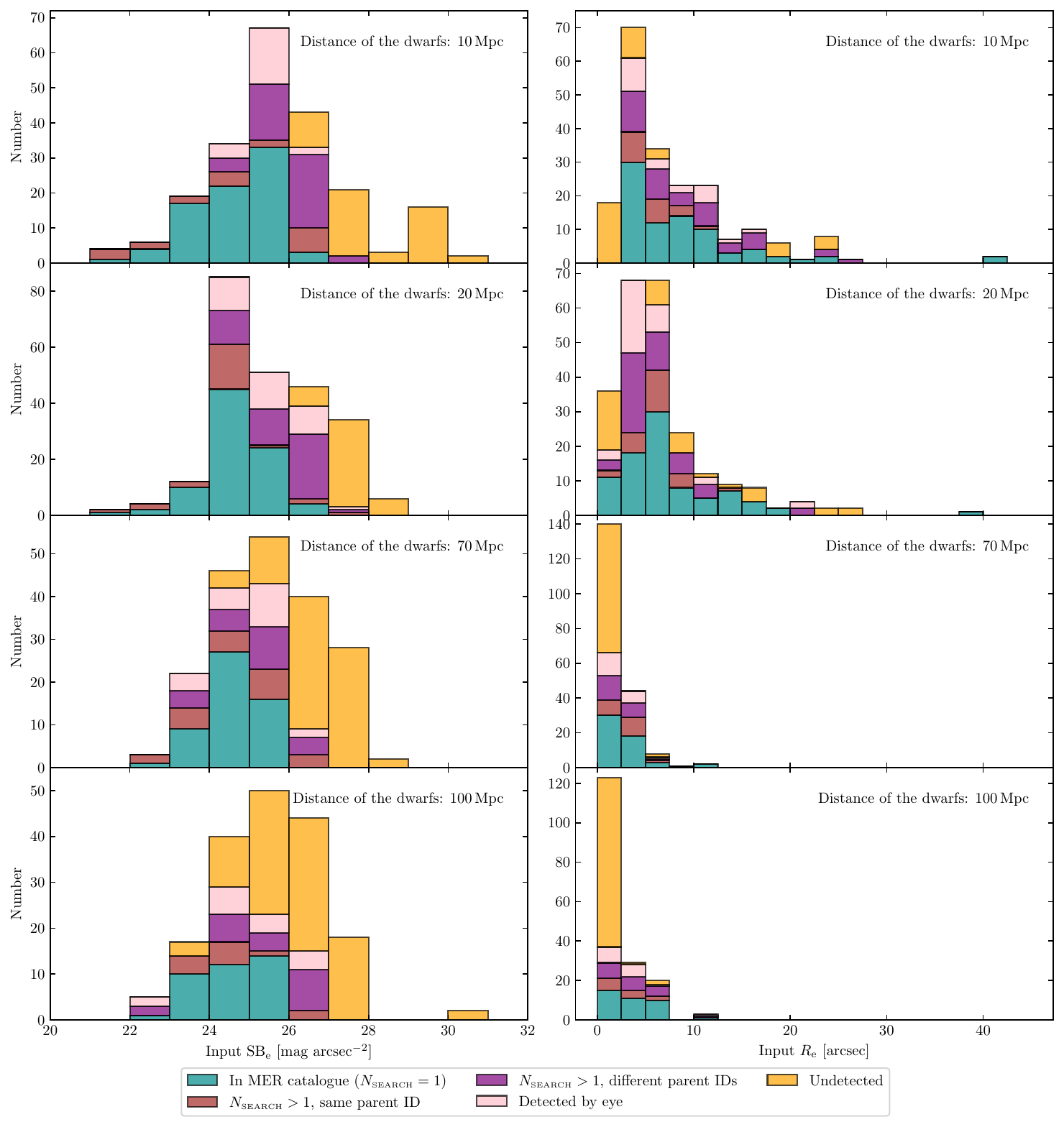}
\vskip -0.3cm
\caption{Histograms of the input ${\rm SB}_{\rm e}$ (left panel) and the input $R_{\rm e}$ (right panel panel) of the dwarfs at 10, 20, 70 and 100\,Mpc, colour-coded according to their detection by eye and in the MER catalogue, with $N_\sfont{SEARCH}$ the number of MER sources found in the search radius $R_{\sfont{SEARCH}}$ used for the cross-match.}
\label{fig:stacked_histo}
\end{figure*}

We provide further detail on the impact of the distance on the dwarf detectability in Fig.\,\ref{fig:stacked_histo} and the associated Tables\,\ref{tab:sb_bins_D} and \ref{tab:re_bins_D}.

\begin{table*}[h!]
\setlength{\tabcolsep}{2.5pt}
\caption{Representative detection statistics for the sample.}
\centering
\renewcommand{\arraystretch}{1.2}
\setlength{\extrarowheight}{2pt}
\begin{tabular}{ccccccc}
\hline \hline
\noalign{\vskip 4pt}
$D$ [Mpc] & ${\rm SB}_{\rm e}$ $[\text{mag\,arcsec}^{-2}]$ & Input dwarfs & In MER catalogue & Single parent ID & Single or multiple parent ID(s) & Detected by eye \\
(1) & (2) & (3) & (4) & (5) & (6) & (7) \\
\noalign{\vskip 4pt}
\hline
\noalign{\vskip 4pt}
10 & 25–26 & $67\,(100\,\%)$ & $33\,(49\,\%)$ & $35\,(52\,\%)$ & $51\,(76\,\%)$ & $67\,(100\,\%)$ \\
20 & 24–25 & $85\,(100\,\%)$ & $45\,(53\,\%)$ & $61\,(72\,\%)$ & $73\,(86\,\%)$ & $85\,(100\,\%)$ \\
70 & 25–26 & $54\,(100\,\%)$ & $16\,(30\,\%)$ & $23\,(43\,\%)$ & $33\,(61\,\%)$ & $43\,(80\,\%)$ \\
100 & 25–26 & $50\,(100\,\%)$ & $14\,(28\,\%)$ & $15\,(30\,\%)$ & $19\,(38\,\%)$ & $23\,(46\,\%)$ \\
\noalign{\vskip 4pt}
\hline \hline
\end{tabular}
\tablefoot{For each distance $D$ (column 1, given in Mpc), we extract the ${\rm SB}_{\rm e}$ bin with the highest number of dwarfs (column 2, given in $\text{mag\,arcsec}^{-2}$). Columns (3) indicates the number of injected dwarfs. The remaining columns are the number of dwarfs that are (4) present in the final MER catalogue as one single object, (5) present in the final MER catalogue as one or several objects sharing the same parent ID, (6) present in the final MER catalogue as one or several objects sharing or not the same parent ID(s), and (7) visually detected, whether or not they are present in the final MER catalogue. As a result, columns (4) to (6) are cumulative. Columns (3) to (7) are given in number and in percent of the injected dwarfs in the corresponding surface brightness bin.}
\label{tab:sb_bins_D}
\end{table*}

\begin{table*}[h!]
\caption{Similar to Table \ref{tab:sb_bins_D}, but with bins in $R_{\rm e}$ (in arcsec) rather than in ${\rm SB}_{\rm e}$.}
\centering
\renewcommand{\arraystretch}{1.2}
\setlength{\extrarowheight}{2pt}
\begin{tabular}{ccccccc}
\hline \hline
\noalign{\vskip 4pt}
$D$ & $R_{\rm e}$ bin & Input dwarfs & In MER catalogue & Single parent ID & Single or multiple parent ID(s) & Detected by eye \\
(1) & (2) & (3) & (4) & (5) & (6) & (7) \\
\noalign{\vskip 4pt}
\hline
\noalign{\vskip 4pt}
10 & 2.5–5 & $70\,(100\,\%)$ & $30\,(43\,\%)$ & $39\,(56\,\%)$ & $51\,(73\,\%)$ & $61\,(87\,\%)$ \\
20 & 2.5–7.5 & $136\,(100\,\%)$ & $48\,(35\,\%)$ & $66\,(49\,\%)$ & $100\,(74\,\%)$ & $129\,(95\,\%)$ \\
70 & 0–2.5 & $140\,(100\,\%)$ & $30\,(21\,\%)$ & $39\,(28\,\%)$ & $53\,(38\,\%)$ & $66\,(47\,\%)$ \\
100 & 0–2.5 & $123\,(100\,\%)$ & $15\,(12\,\%)$ & $21\,(17\,\%)$ & $29\,(24\,\%)$ & $37\,(30\,\%)$ \\
\noalign{\vskip 4pt}
\hline \hline
\end{tabular}
\label{tab:re_bins_D}
\end{table*}

\section{\label{Detectability_nuclei}Details on the visual inspection}

Here, we detail the results of our visual inspection in each dwarf cutout for the three types of image (NOBG, VIS BGSUB, and final MER mosaics). The motivations for this study are as follows:
\begin{itemize}
\item To identify and exclude cases where source matching between the input dwarf catalogue and the final MER catalogue is unreliable due to contamination or confusion with background objects. Such cases are removed from the corrected detection statistics discussed in the main text. The results before that correction are available in Appendix \ref{App_non_corrected_det}.
\item Evaluating the impact of the background subtraction on the detection of dwarf galaxies, with or without a nuclear star cluster. We detail below the results of this study.
\end{itemize}

\begin{figure*}[h]
\centering
\includegraphics[width=\linewidth]{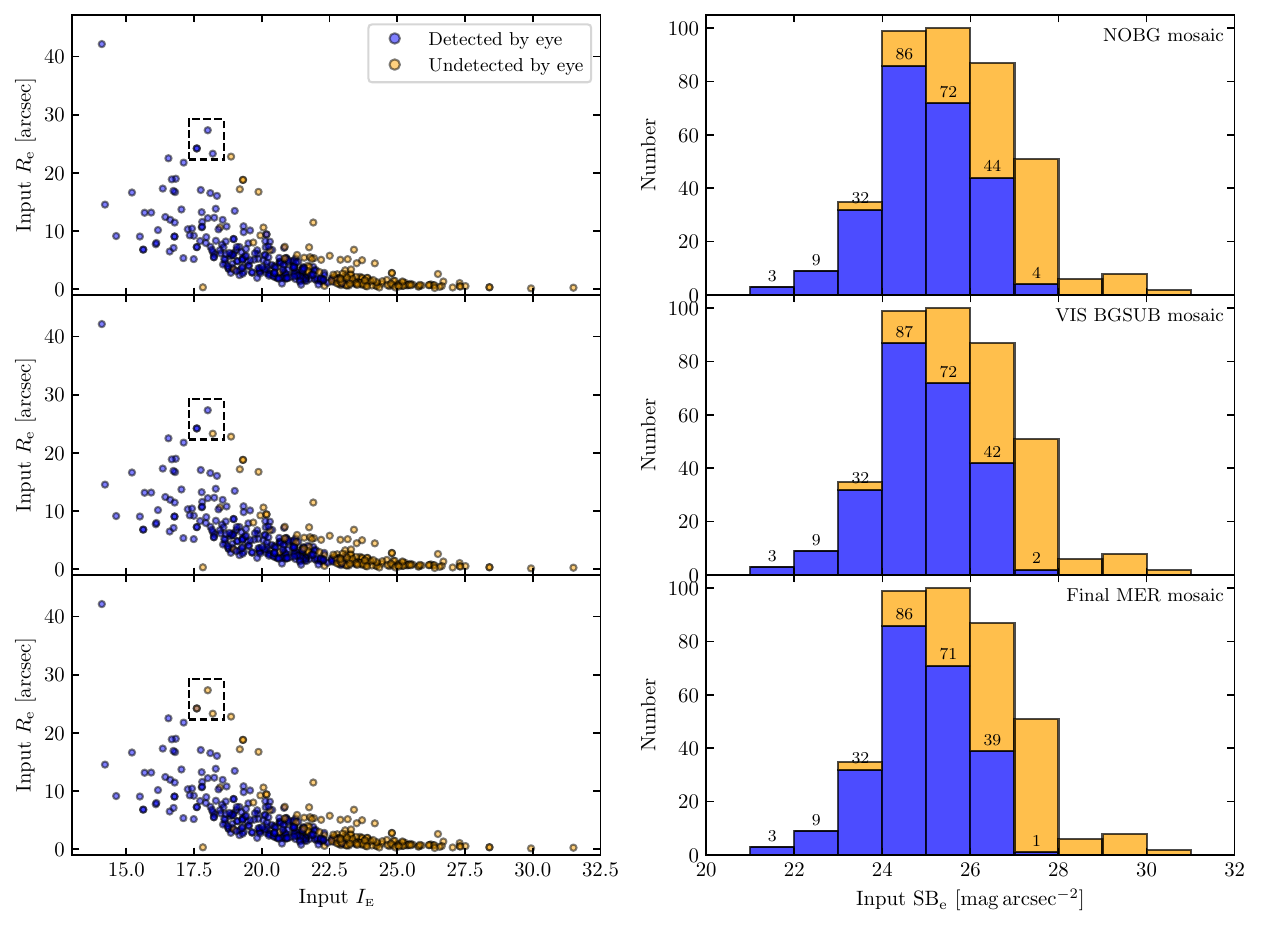}
\caption{\label{fig:detection}Input $R_{\rm e}$ as a function of the input total magnitude in $\IE$ (left panel) and the corresponding histogram of the input ${\rm SB}_{\rm e}$ (right panel), colour-coded according to their detection or non-detection by eye in the three types of products: NOBG, VIS BGSUB, and the final MER mosaic subtracted by VIS and MER backgrounds. Only the non-nucleated dwarfs, at all distances, have been used in this analysis.  We labelled the number of dwarfs that are visually detected in each ${\rm SB}_{\rm e}$ bin. The dashed rectangle shows examples of objects that were detected differently with the various background treatments.}
\end{figure*}

Figure\,\ref{fig:detection} compiles the results of the visual detection for galaxies at all distances and demonstrates the visual detectability of each dwarf according to their input parameters. Nucleated dwarfs were excluded from this test, to ensure that we are not impacted by the detectability of the nucleus alone. The left panel includes parameter space plots (the input $R_{\rm e}$ as a function of the input $\IE$) which are colour-coded according to the visual detectability of the dwarf for the three types of products (NOBG, VIS BGSUB, and final MER mosaics).
We then construct the right panel of Fig.\,\ref{fig:detection}, i.e. the histograms of ${\rm SB}_{\rm e}$ for the three products, using the detection information already employed in the previous plot.

\begin{figure*}[h]
\centering
\includegraphics[width=\linewidth]{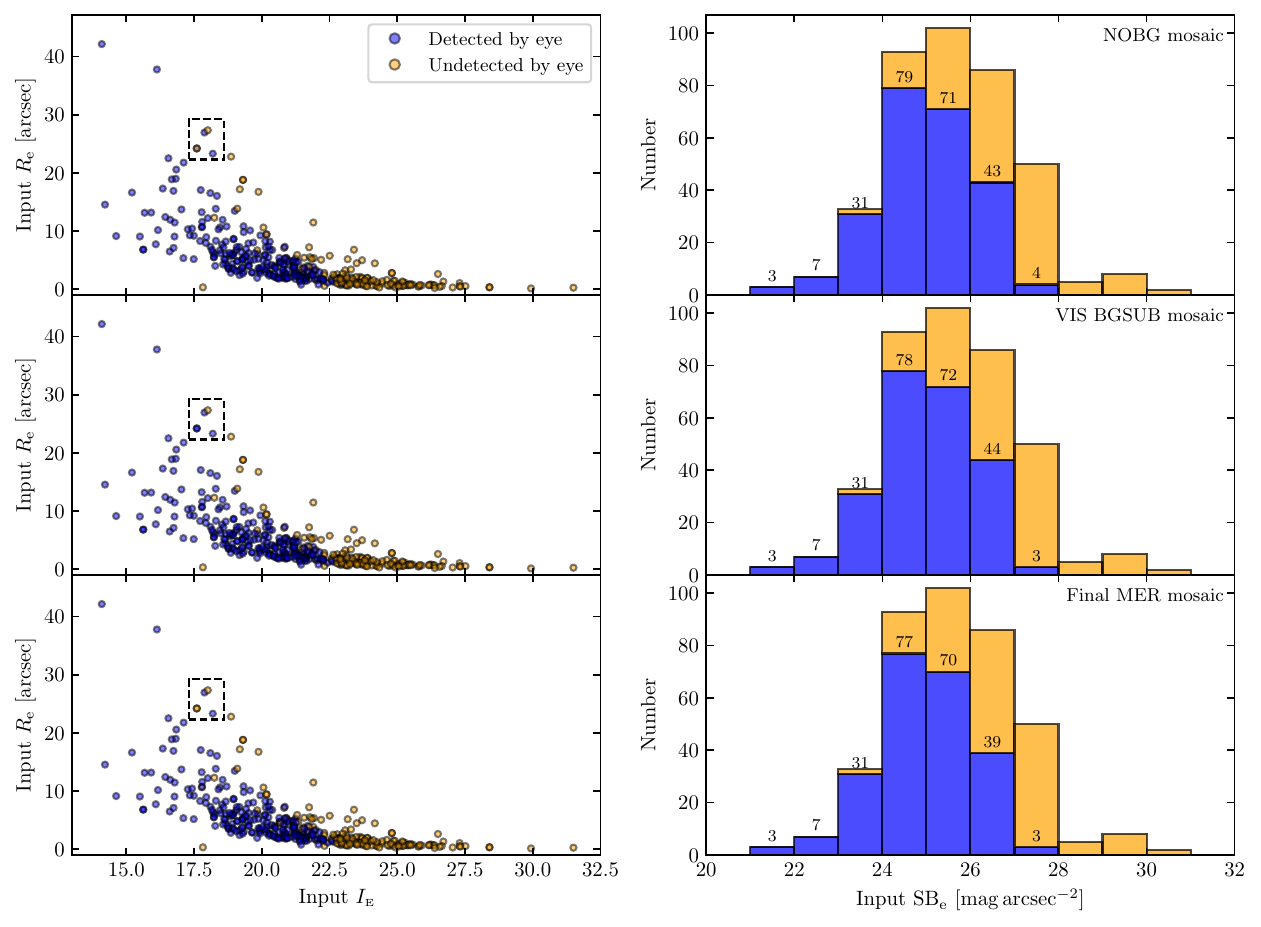}
\caption{\label{fig:detection_nuclei}Identical to Fig.\,\ref{fig:detection}, but for nucleated dwarf galaxies.}
\end{figure*}

Finally, we repeat the above analysis for the nucleated dwarfs, as shown in Fig.\,\ref{fig:detection_nuclei}. When we compare with Fig.\,\ref{fig:detection}, we note that the presence of a nucleus has a slight impact on dwarf detection, but does not significantly alter the distribution within the parameter space or the ${\rm SB}_{\rm e}$ histogram. For a few faint dwarfs, it seems to facilitate their identification (examples are provided in the highlighted region of the left panel, where some nucleated dwarfs remain detected after the background subtraction whereas the non-nucleated dwarfs are not). This is the reason why fewer dwarfs are lost among the background subtractions in the case of nucleated dwarfs (five lost with $26\,\text{mag\,arcsec}^{-2}<{\rm SB}_{\rm e}<28\,\text{mag\,arcsec}^{-2}$, compared to eight for non-nucleated dwarfs).

We are now interested in the impact of the background subtractions. We first focus on comparing the rows of Figs.\,\ref{fig:detection} and \ref{fig:detection_nuclei} that pertain to the final MER mosaic and the VIS BGSUB mosaic, in order to probe the effect of the second, local, MER background subtraction. It has very little impact on the parameter space plot and on the global shape of the ${\rm SB}_{\rm e}$ histogram. Between the two background subtractions, only a small number (14, or $2\,\%)$ of the injected dwarfs, highlighted in the dashed box in the parameter space plots, are lost above ${\rm SB}_{\rm e} = 24\,\text{mag\,arcsec}^{-2}$, showing that the subtraction slightly affects only the LSB regime, while remarkably preserving the vast majority of detectable dwarfs.

We then focus on comparing the first and second rows of Figs.\,\ref{fig:detection} and \ref{fig:detection_nuclei} that pertain to the VIS BGSUB and NOBG mosaics, in order to probe the effect of the first VIS background subtraction. Our conclusions on the dwarf detectability are similar to those reported for the MER second local background subtraction, i.e. we observe very little impact on dwarf detection rates, losing a very small number of dwarfs above ${\rm SB}_{\rm e} = 24\,\text{mag\,arcsec}^{-2}$ (7 dwarfs or $1\,\%$ of the injected dwarfs).

\FloatBarrier

\end{document}